\RequirePackage{fix-cm}
\documentclass[twocolumn,epjc3]{svjour3}  
\smartqed  
\RequirePackage{graphicx}

\usepackage{multirow, makecell} 
\usepackage{comment} 
\usepackage{amsmath, mathtools}
 
\usepackage{url}

\usepackage{breakurl}
\usepackage[breaklinks, hidelinks]{hyperref}

\usepackage[
backend=biber,
style=phys,
sorting=none,
maxcitenames=2,
maxbibnames=2,
]{biblatex}
\usepackage[left]{lineno}

\newcommand{\insertTextInFigure}[4]{%
    \begin{picture}(0,0)
        \put(#1,#2){\fontfamily{\sfdefault}\fontsize{#4}{#4}\selectfont\text{#3}}
    \end{picture}
}

\journalname{Eur. Phys. J. C}

\begin{document}

\title{Scintillation Light in SBND: Simulation, Reconstruction, and Expected Performance of the Photon Detection System
}

\authorrunning{The SBND Collaboration}

\author{ The SBND Collaboration\thanksref{e1}\\\\
P.~Abratenko~\thanksref{addr38}\and
R.~Acciarri~\thanksref{addr15}\and
C.~Adams~\thanksref{addr1}\and
L.~Aliaga-Soplin~\thanksref{addr37}\and
O.~Alterkait~\thanksref{addr38}\and
R.~Alvarez-Garrote~\thanksref{addr9}\and
C.~Andreopoulos~\thanksref{addr21}\and
A.~Antonakis~\thanksref{addr4}\and
L.~Arellano~\thanksref{addr24}\and
J.~Asaadi~\thanksref{addr37}
% ...
}
\institute{
Argonne National Laboratory, Lemont, IL 60439, USA \label{addr1}
% ...
\and\pagebreak[0] University of California, Santa Barbara CA, 93106, USA \label{addr4}
% ...
\and\pagebreak[0] CIEMAT, Centro de Investigaciones Energ\'{e}ticas, Medioambientales y Tecnol\'{o}gicas, Madrid E-28040, Spain \label{addr9}
% ...
\and\pagebreak[0] Fermi National Accelerator Laboratory, Batavia, IL 60510, USA \label{addr15}
% ...
\and\pagebreak[0] University of Liverpool, Liverpool L69 7ZE, United Kingdom \label{addr21}
% ...
\and\pagebreak[0] University of Manchester, Manchester M13 9PL, United Kingdom \label{addr24}
% ...
\and\pagebreak[0] University of Texas at Arlington, TX 76019, USA \label{addr37}
\and\pagebreak[0] Tufts University, Medford, MA, 02155, USA \label{addr38}
% ...
}
\thankstext{e1}{sbnd\_info@fnal.gov}

\onecolumn
\maketitle
\twocolumn
\sloppy

\begin{abstract}
SBND is the near detector of the Short-Baseline Neutrino program at Fermilab. Its location near to the Booster Neutrino Beam source and relatively large mass will allow the study of neutrino interactions on argon with unprecedented statistics. This paper describes the expected performance of the SBND photon detection system, using a simulated sample of beam neutrinos and cosmogenic particles. Its design is a dual readout concept combining a system of 120 photomultiplier tubes, used for triggering, with a system of 192 X-ARAPUCA devices, located behind the anode wire planes. Furthermore, covering the cathode plane with highly-reflective panels coated with a wavelength-shifting compound recovers part of the light emitted towards the cathode, where no optical detectors exist. We show how this new design provides a high light yield (LY) and a more uniform detection efficiency, an excellent timing resolution and an independent 3D-position reconstruction using only the scintillation light. Finally, the whole reconstruction chain is applied to recover the temporal structure of the beam spill, which is resolved with a resolution on the order of nanoseconds.
\end{abstract} 

\section{Introduction}
\label{intro}
The Short-Baseline Near Detector (SBND) is one of the detectors making up the Short-Baseline Neutrino (SBN) Program at Fermilab~\cite{Machado_2019,acciarri2015proposal}. The main goals of this program are to address the existence of eV-scale sterile neutrinos, to study neutrino-nucleus interactions with high statistics in the GeV energy range, and to contribute to the advancement of liquid argon detector technology. 
SBND will measure neutrino interactions from the Booster Neutrino Beam (BNB), with an energy spectrum peaking at $\sim$700\,MeV. Located at a distance of 110\,m from the start of the beamline, SBND will characterise the BNB flux before significant flavour oscillations occur, recording millions of neutrino interactions per year. This event rate enables SBND to perform neutrino-argon cross-section measurements with the highest statistics to date for many processes, as well as search for rare events both in the Standard Model and beyond. The role of SBND will therefore be key not only for the SBN physics program, but also for future long-baseline neutrino experiments, such as the Deep Underground Neutrino Experiment (DUNE)~\cite{DUNE:2020lwj}.

\begin{figure}[t]
    \centering
         \includegraphics[width=1.\linewidth,  angle=0]{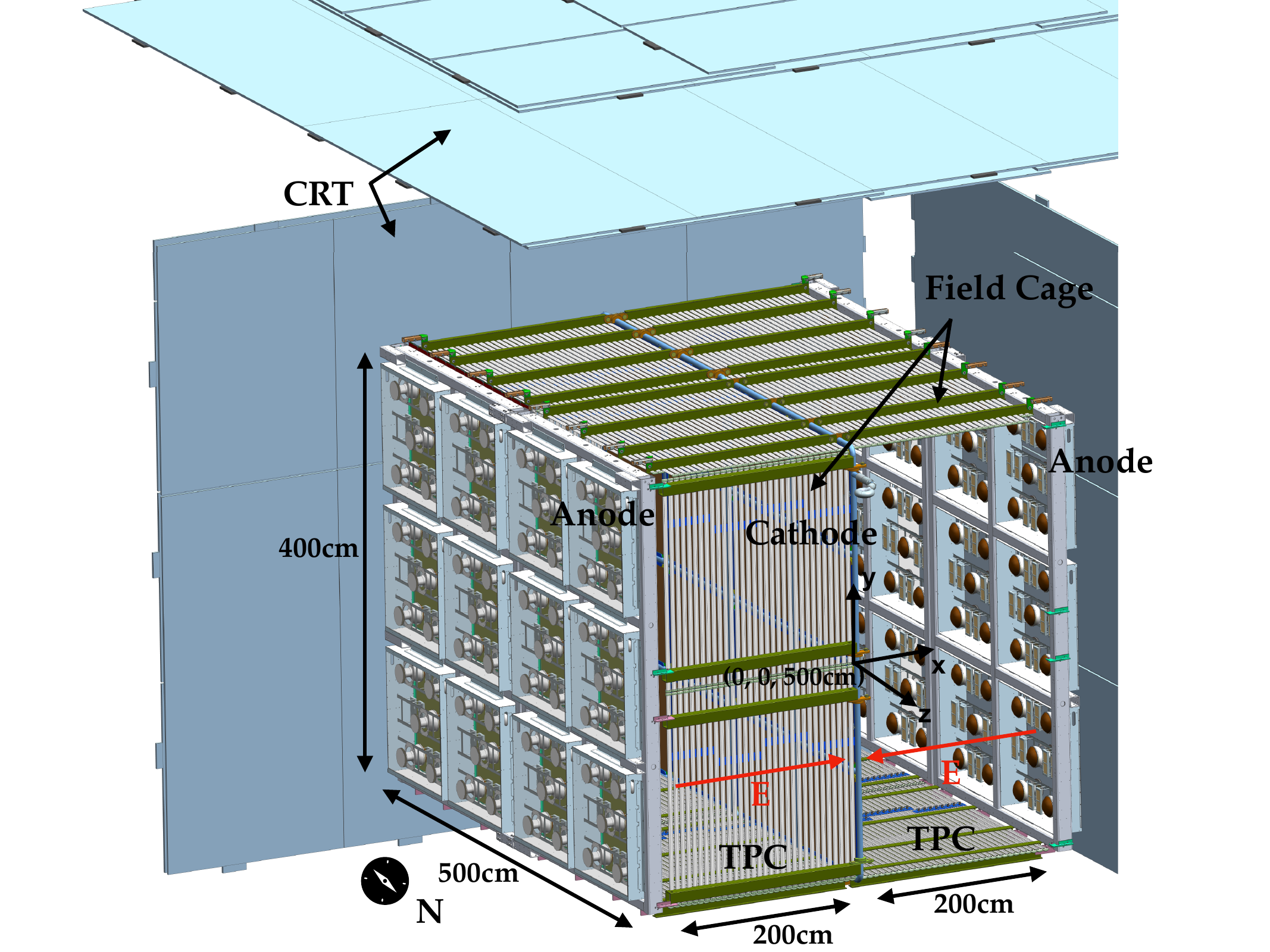}
     \caption{Drawing showing the SBND detector. The cryostat, part of the vertical field cage and some CRT panels (facing the north, east and bottom TPC walls) have been removed to reveal the active volume. The neutrino beam enters the detector from the south side.}\label{fig:SBNDdetector}
\end{figure}

SBND is a liquid argon time projection chamber (LArTPC)~\cite{Rubbia:1977zz} with an active mass of 112\,tons and size 400\,cm (X-drift) $\times$ 400\,cm (Y-height) $\times$ 500\,cm (Z-length). It is divided into two drift regions of 200\,cm each, defined by a central cathode and two anodes surrounded by the field cage structure, as shown in Figure~\ref{fig:SBNDdetector}. Each anode integrates three wire planes at different orientations (0$^\circ$, +60$^\circ$, -60$^\circ$ to the vertical) with 3\,mm spacing between wires within a plane and between the three planes~\cite{SBND:2020scp}. These wires record the signals due to the drifting electrons from the ionisation produced by charged particles passing through the detector. The drift coordinate is perpendicular to the beam in the horizontal direction, and the maximum drift time corresponds to about 1.3\,ms for the nominal electric field of 500\,V/cm. Since SBND is located near the surface, the detector is surrounded on all sides by scintillator strip planes that serve as a cosmic ray tagger (CRT) and provide the trajectory and timing of particles (mainly muons) that pass through the detector walls.

Together with ionisation, charged particle interactions produce excited argon molecules which emit scintillation light as they dissociate to their ground atomic state. These photons, which traverse the detector on a time scale of nanoseconds compared to $\mathcal{O}(\rm ms)$ electron drift times, are collected by photon detector (PD) devices. The detection of these photons is typically used to determine the start time of the interactions. Although in these detectors the deposited energy is divided between the generation of electrons and photons, past LArTPC neutrino experiments have primarily focused on exploiting the information provided by the charge using the light information in a more limited way. However, the trend is changing in newer LArTPC designs.

The complementarity between the charge and light signals generated in a LArTPC makes it clear that an advanced photon detection system (PDS) offers the potential to significantly improve the performance of the experiment. The LArIAT experiment has demonstrated light-augmented calorimetry for low-energy electrons in a small LArTPC~\cite{PhysRevD.101.012010}, but it remains to be demonstrated in large detectors. Other benefits will include light-enhanced particle identification, on-the-fly position reconstruction and improved timing resolution. This will not only maximise SBND's performance, but also presents an R\&D opportunity for light detection in liquid argon, with implications for future experiments such as DUNE. While the scope of this article is restricted to scintillation light signals, in future work we will study the performance when correlating the light signals with the ionisation objects from the TPC and CRT systems.

This article is organised as follows: in section~\ref{sec:SBND-PDS} we describe the novel design of the PDS in SBND. Section~\ref{sec:lightgeneration} illustrates the approach followed for the simulation of the light signals, from generation to detection of the photons. In section~\ref{sec:detresponsesim} we explain all the detector effects related to the processing of the light signals. The reconstruction strategy followed in this work is described in section~\ref{sec:lightreco}. Section~\ref{sec:reco_effficiency} shows the reconstruction efficiency of the light signals while in section~\ref{sec:reco_performance} we evaluate the performance of our approach for the calorimetric, spatial, and timing resolution. Finally, as a case study, in section~\ref{sec:BNBreconstruction} we apply the full simulation and reconstruction chain described above to assess the ability of SBND to resolve the time structure of the BNB, using only the light detection system.
 
\section{Photon detection system design}\label{sec:SBND-PDS}

Scintillation photons in argon are emitted in the vacuum ultraviolet (VUV) region of the electromagnetic spectrum with the maximum intensity peak at 128\,nm and a FWHM of about 6\,nm~\cite{Doke:1981eac,PhysRevB.27.5279}. These VUV photons are absorbed by most materials, making them undetectable by standard devices. To make the detection of scintillation light of liquid argon possible, parts of the detector (e.g. the windows of the optical sensors) are typically coated with wavelength-shifter (WLS) compounds which re-emit light in the wavelength range where it can be detected more efficiently.

The PDS in SBND integrates two different technologies: (i) a system of 120 cryogenic 8-inch diameter optical photomultipliers (PMTs), Hamamatsu R5912-MOD model~\cite{R5912datasheet}, and (ii) a system of 192 X-ARAPUCA devices\footnote{Similar devices are planned for the phase\,I DUNE Far Detectors \cite{DUNE:2020txw}.}, based on a photon trap concept~\cite{arapucas}. The X-ARAPUCA confines the wavelength-shifted photons inside a highly reflective box where silicon photomultipliers (SiPMs) capture them, effectively increasing their collection area. Both systems are located behind the wire planes of each TPC. The PMT system is read out using 500\,MHz sampling CAEN VX1730SB digitiser modules. The X-ARAPUCAs are read out using an amplifier similar to the DUNE model~\cite{Brizzolari:2022fzb} but located outside the cryostat, and digitised using 62.5\,MHz sampling CAEN V1740B modules. One of the main goals for the X-ARAPUCA system in SBND is to serve as R\&D for this new technology and to demonstrate its operation in a neutrino beam.  

In order to maximise the number of detected photons in SBND, the TPC cathode surface has been covered with reflective foils coated with te\-tra\-phe\-nyl-bu\-ta\-di\-ene (TPB)\footnote{The coating was carried out by an evaporation process reaching an average mass density of $\sim$300\,\textmu g/cm$^2$~\cite{phdthesisVincent}.}~\cite{refTPB}. The foils are held in place using a mesh, with a 79\% transmittance, that covers the full foils surface. The use of such foils is common in dark matter experiments~\cite{Acciarri:2010zz,Regenfus:2009fu}. The LArIAT detector~\cite{LArIAT:2019kzd} was the first to successfully implement them in a single-phase LArTPC using test-beam data, but SBND is the first to do so to study neutrinos, and at an unprecedented scale. This design allows us to recover part of the light emitted in the opposite direction to the plane where the PDs are located that would otherwise be lost. Therefore, the PDS of SBND is sensitive to two different light components: (i) a {\it direct} component, the photons arriving with VUV wavelengths to the detector windows where they are wavelength-shifted to the visible and then detected, and (ii) a {\it re-emitted}/{\it reflected} component, the photons arriving with visible wavelengths to the detector windows and then detected. Note that the component called here direct is the one detected by other single-phase large LArTPC detectors (e.g. MicroBooNE~\cite{MBooNEProposal}, ProtoDUNE-SP~\cite{DUNE:2020cqd} or ICARUS~\cite{IcarusFirstOp}).

In SBND only the PMT system is used to build trigger signals, and is therefore considered as the primary light detection system. The 120 PMTs are organised into two arrays of 60 PMTs each, installed in the two optically-isolated TPC volumes. In each TPC, the windows of 48 PMTs are TPB-coated and thus sensitive to both direct and reflected light components; the remaining 12 PMTs per TPC are left uncoated and thus only sensitive to reflected light. This coated/uncoated configuration is chosen to ensure high light collection while maintaining the ability to distinguish between the different light components, enabling the extension of the use of scintillation light in event reconstruction, as will be discussed in detail in section~\ref{sec:reco_performance}.

Similarly, the X-ARAPUCA system is made up of 96 devices per TPC. Half of them are coated with para-terphenyl (pTP)~\cite{refpTP} to allow the detection of direct photons, while reflected photons are detected by both uncoated and coated X-ARAPUCAs. In order to trap the photons, dichroic filter windows (147\,cm$^2$ in size, with nominal cutoffs at 400\,nm and 450\,nm for coated and uncoated units, respectively) paired with WLS bars are used inside the X-ARAPUCAs. The actual cutoff wavelength of the dichroic filters shows a dependence on the angle of incidence of light, the smaller the angle the larger the cutoff. As a result, the coated units whose transmittance range is selected for the pTP emission let some visible light through.

\begin{figure}[t]
    \centering
         \includegraphics[width=1.\linewidth, angle =0]{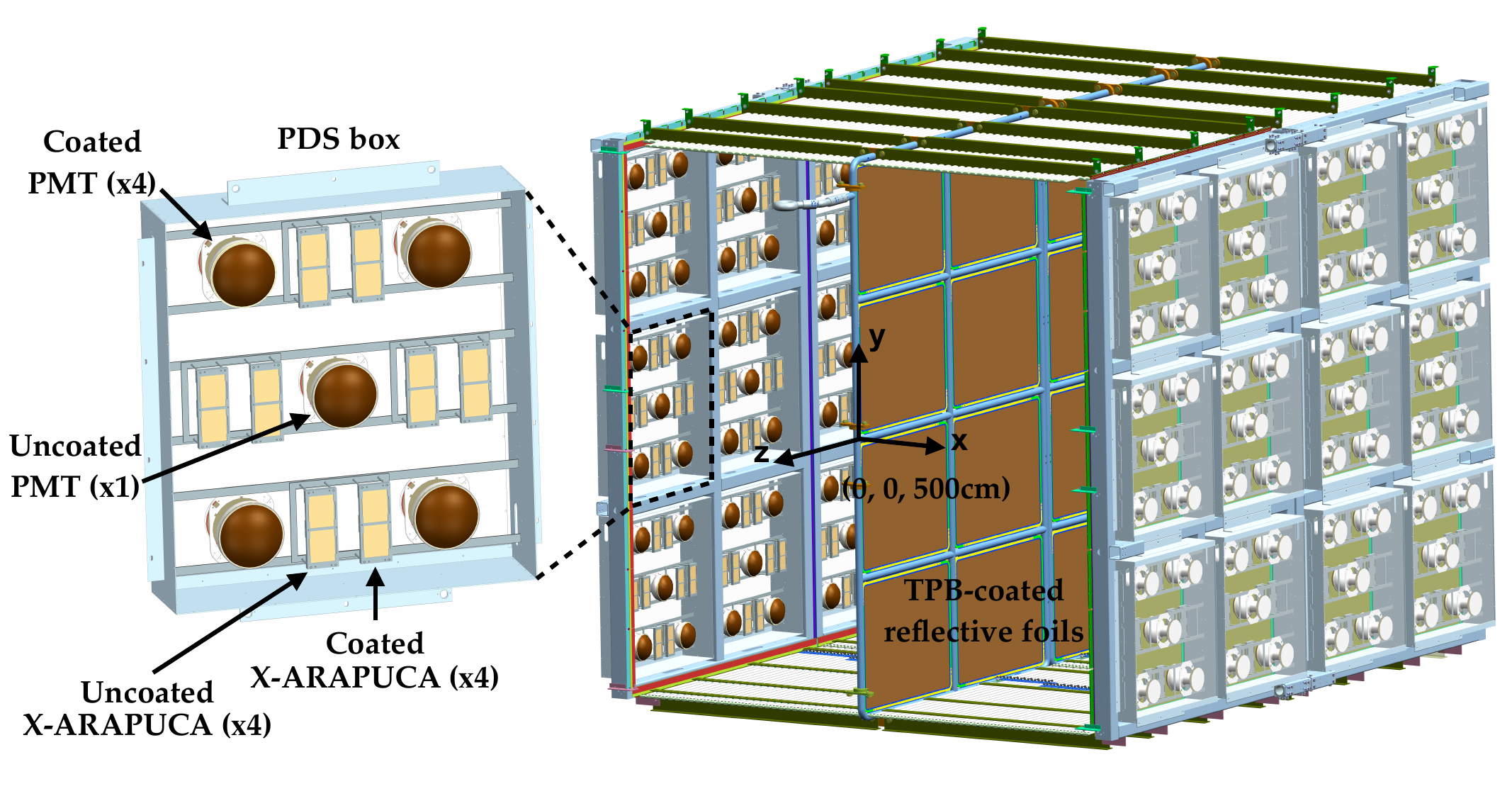}
     \caption{PMT and X-ARAPUCA arrangement in a PDS-box (left), together with a view of SBND's photon detection system (right), including the definition of the coordinate system we use in our work.}
    \label{fig:PDS}      
\end{figure}

Both PMTs and X-ARAPUCAs are grouped into units that we call PDS boxes, with 5 PMTs and 8 X-ARAPUCAs per box as illustrated in Figure~\ref{fig:PDS}. Within each of the two TPC volumes, 12 PDS boxes are arranged in a 4$\times$3 array directly behind the wires. In order to shield the wires from PDS pulses, a metallic mesh with 85\% transmittance is installed between the TPC wire planes and the PDS units.

With this design, where different active (PMTs and X-ARAPUCAs) and passive (WLS reflective foils on the cathode) optical components are combined, the PDS of SBND is the most sophisticated ever installed in a LArTPC. This represents a major R\&D opportunity that contributes to the further development of this technology and helps build the expertise of the worldwide neutrino physics community working on future experiments such as DUNE. 

\section{Scintillation light simulation}\label{sec:lightgeneration}

 The simulation and reconstruction of events in SBND is carried out using the LArSoft package~\cite{larsoft1,Snider:2017wjd}. The particle tracking in LArSoft is done using Geant4~\cite{bib:geant4}. For each energy deposition within the active volume of the detector, LArSoft generates a certain correlated number of electrons and photons depending on the value of the ionisation density and electric field~\cite{PhysRevD.101.012010,LArQL}. The sum of both contributions is proportional to the total deposited energy. Charge and light separately can also be used to estimate the deposited energy, but in these cases it is necessary to apply non-trivial, model-dependent corrections to account for charge lost through recombination effects~\cite{AMORUSO2004275,RAcciarri_2013}. After being generated, simulated scintillation photons and ionisation electrons need to be transported from their production positions to the readout sensitive channels. Henceforth we will only focus on the simulation of the scintillation light (an overview of the charge treatment can be found in~\cite{MicroBooNE:2018swd,MicroBooNE:2018vro}).

Scintillation photons can undergo different physical processes as they propagate: Rayleigh scattering, reflections and refractions in the detector material boundaries, absorptions and wavelength shifting. A full Geant4 simulation that tracks every single optical photon taking into account all these processes is available in LArSoft. However, liquid argon emits $\mathcal{O}$(20000) photons per MeV of deposited energy at 500\,V/cm (and twice as much in the absence of electric field)~\cite{DOKE1990617,Doke_2002}. This makes the full Geant4 simulation very CPU-intensive and prohibitively slow, especially for large detector sizes and $\mathcal{O}$(GeV) energy depositions as is the case for BNB neutrinos in SBND. This makes our scintillation light simulation computationally challenging and alternative methods, commonly known as {\it fast optical models}, need to be considered. We will next introduce the models that we use within LArSoft to predict the number of photons arriving to the PDs and their arrival time distributions.

\subsection{Number of photons detected}
In SBND we use the semi-analytic model described in~\cite{PredTransportScintLight} to simulate the transport effects in scintillation light signals. This approach makes use of the isotropic emission of the scintillation photons to calculate the geometrical aperture of each PD relative to each scintillation point on-the-fly. Corrections are then applied to the photon transport to account for Rayleigh scattering and border effects due to the finite size of the detector.
The model also includes an extension to accurately simulate the reflected component of the light signals in SBND. This approach, which does not scale with the size of the detector, is drastically faster (more than $\times$10) than the full Geant4 optical simulation.

One limitation of the semi-analytic method is that it can only be defined for the active volume due to its geometric approach. However, light generated outside of the TPC (especially behind the wire planes) might have a non negligible contribution to the signals. For example, though the light produced in the non-active volume can be considered as a second order contribution, it can play a crucial role for trigger efficiency studies, where crossing cosmic tracks behind the PDS can lead to fake triggers. To simulate the light generated outside the active volume in SBND we use the optical library model. It is available in LArSoft and has been the default light simulation mode pioneered by the LArIAT~\cite{PhysRevD.101.012010} and MicroBooNE~\cite{MicroBooNE:2021zul} experiments. In this approach, the fraction of incident photons for each PD and detector-location pair is computed once using the Geant4 optical simulation and stored in a library file, so that it can be read later in subsequent detector simulations. SBND therefore uses a hybrid model to simulate scintillation light that takes advantage of both the semi-analytic and the optical library approaches.
\begin{figure*}[t]
    \includegraphics[width=.49\textwidth]{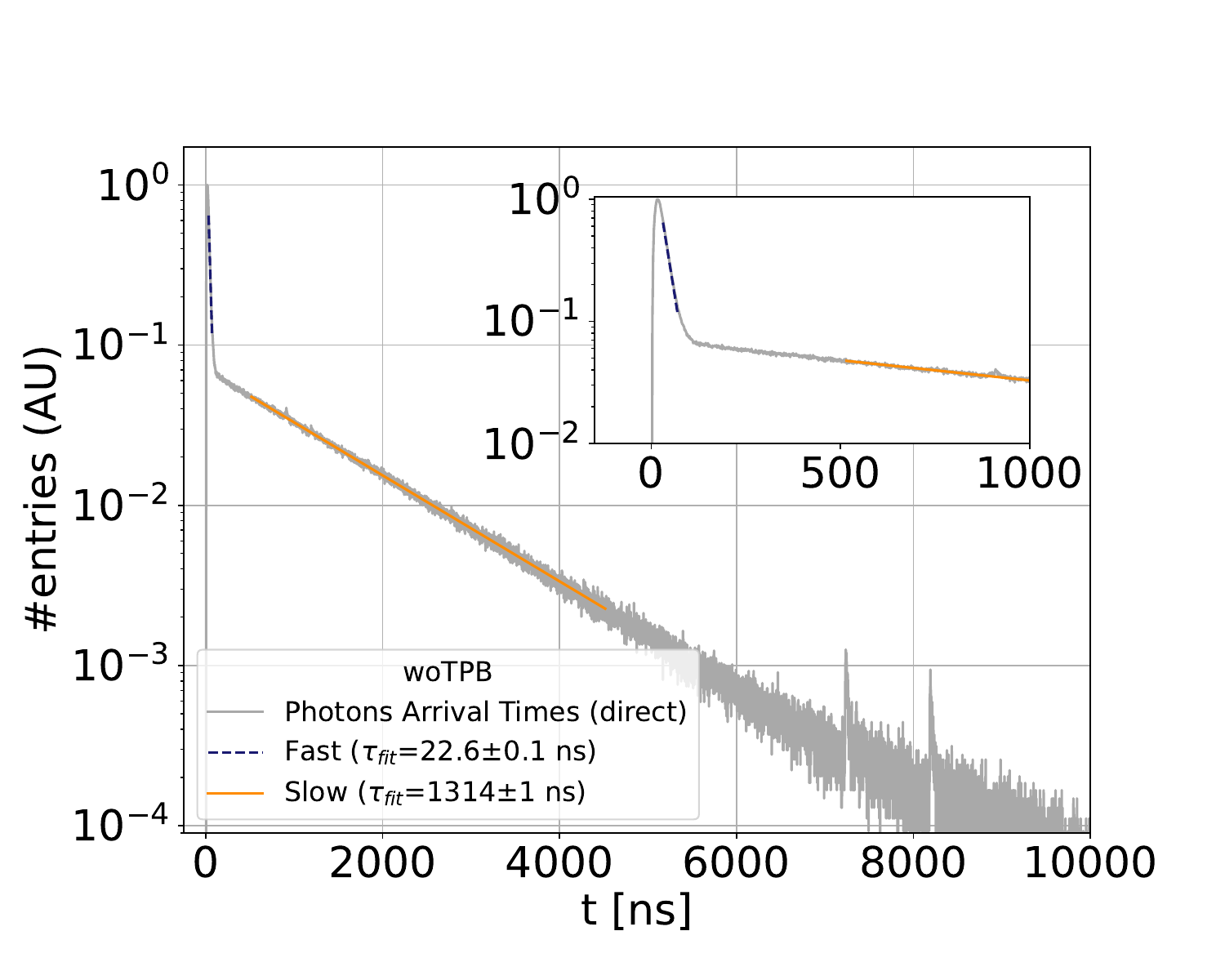}
    \includegraphics[width=.49\linewidth]{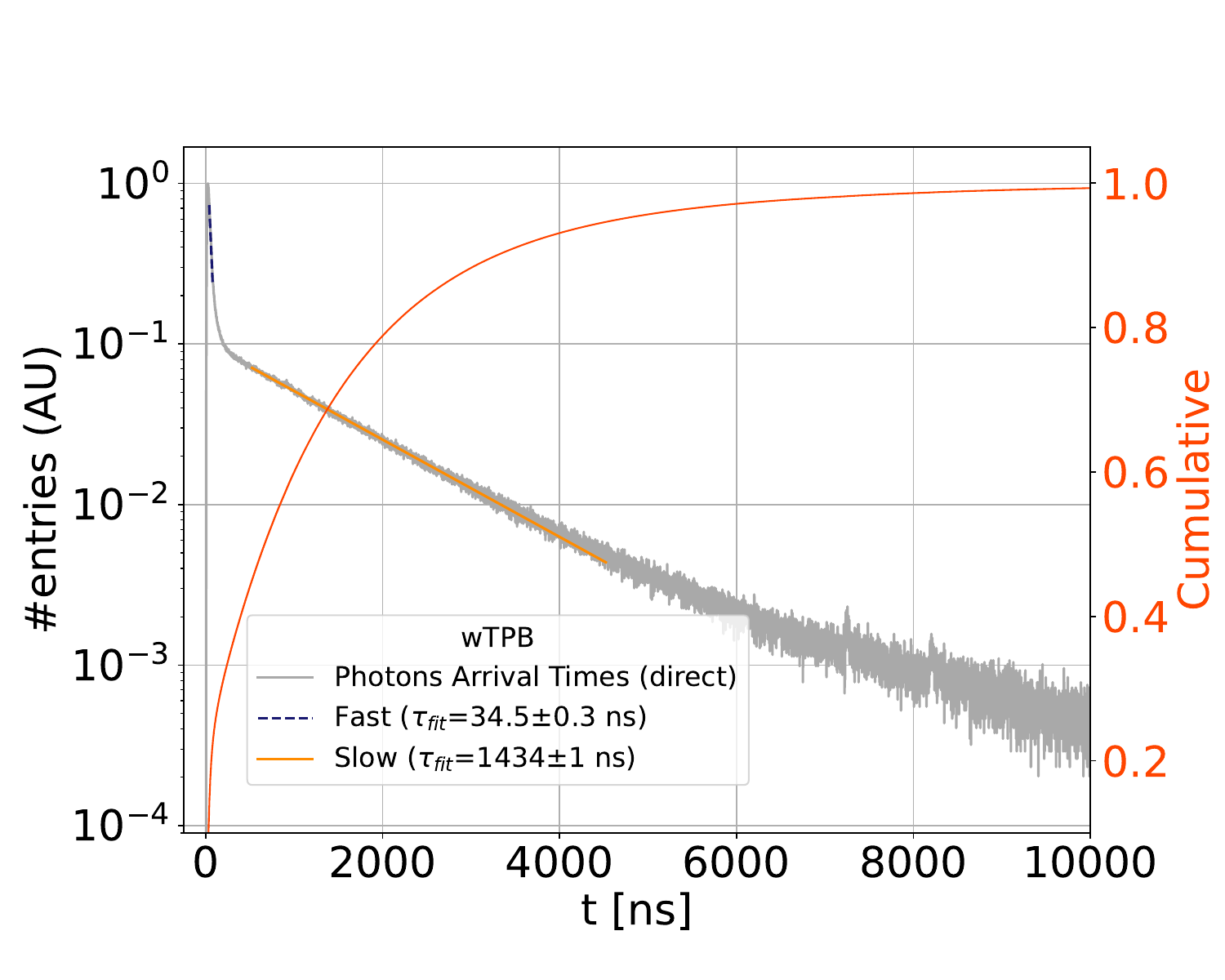}
    \caption{Left: Scintillation photon arrival time distribution for a BNB neutrino sample (1,000 events). Both emission and propagation effects are simulated. Right: Average photon arrival time distribution accounting for TPB emission time. } 
    \label{fig:scintemission}
\end{figure*}

\subsection{Arrival time distributions}

The other key component needed to have a complete light simulation is the time structure of the scintillation signals. In all measurements the overall scintillation light emission in liquid argon exhibits a double exponential decay characterised by two very different times, commonly known as {\it fast} and {\it slow} components~\cite{PhysRevB.27.5279,PhysRevB.20.3486}. The value used in our simulations for the fast component is $\tau_{\text{fast}}$ = 6\,ns. The slow component $\tau_{\text{slow}}$ is less well known with measurements ranging from 1000\,ns to 1700\,ns. Moreover, the existence of an intermediate third component (with a time constant of about 100\,ns), first reported by various experiments~\cite{Heindl_2010,Acciarri_2010_thirdcomp}, has been proposed. However, recent studies \cite{Segreto_2015} have shown that these discrepancies on the value of $\tau_{\text{slow}}$ could be related to the intrinsic emission time of the wavelength shifters needed to enable the VUV photons detection. Dedicated measurements of light signals without WLS\footnote{VUV photons are directly detected by MgF$_2$-window PMTs.} reported a slow decay time of $\tau_{\text{slow}}$ = 1300\,ns \cite{Heindl_2010}, which is the value used in SBND simulations.

In a detector the size of SBND, photon trajectories will be affected at first order by Rayleigh scattering, and at second order by reflections from solid surfaces in the detector volumes. These effects will act cumulatively to lengthen and broaden the time distribution of photons arriving at each PD, especially for detectors whose size is comparable to the Rayleigh scattering length in liquid argon of $\sim$100\,cm~\cite{MBabicz2020LightPI}.
To account for these transport delays in our simulations, we use the semi-analytic model's method to estimate the photon propagation time~\cite{PredTransportScintLight}.

Figure~\ref{fig:scintemission}-Left shows the averaged photon arrival time distribution of the direct light component generated by a simulated sample of BNB neutrino events interacting in SBND. An exponential fit to the fast (blue/dashed-line) and slow (orange/solid-line) components has been performed. As expected, photon transport effects are more pronounced for the fast scintillation component, producing a relative deviation of $\sim400\%$ with respect to the emission lifetime. In comparison, the slow component is largely unaffected, with the expected distribution smeared by only $\sim$1\%.

As mentioned above, the wavelength shifting process is not instantaneous. According to reference~\cite{Segreto_2015}, the TPB time response can be described by a four-exponential function with time constants (abundances) given by: $\tau_1$ $<$ 10\,ns (60\%), $\tau_2$ = 49\,ns (30\%), $\tau_3$ = 3550\,ns (8\%), and $\tau_4$ = 309\,ns (2\%). The TPB effectively changes the time structure of the light signals. In SBND this delay will be present in all PMT light signals. Figure~\ref{fig:scintemission}-Right shows the photon arrival time distribution for the same neutrino sample used in Figure~\ref{fig:scintemission}-Left after including the TPB re-emission time. It can be seen how the effective lifetimes become larger ($\sim$50\% and $\sim$10\% for the fast and slow components respectively) with respect to the averaged emission + propagation time constants. In order to quantify the scintillation signal time density for a given interaction in the LAr, Figure~\ref{fig:scintemission}-Right also shows the cumulative distribution (right vertical-axis). It is interesting to notice that about 99\% of the scintillation photons are expected to reach the PDs within the first $\sim$8\,\textmu s. This result helps inform the length of the integration window set in the reconstruction.

\begin{table}
\begin{center}
    \begin{tabular}{|c|c|c|}
    \hline
        \textbf{\makecell{Light \\ Component}} &  \textbf{\makecell{PMT \\PDE [\%]}} & \makecell{\textbf{X-ARAPUCA} \\ \textbf{PDE} [\%]~\cite{SBND_Internal}} \\
         \hline
        Coated / VUV & 12\%~\cite{Bonesini_2018} & 2.19\%\\
        Coated / Visible & 17\%~\cite{SBND_Internal} & 0.43\%\\
        Uncoated / VUV & 0\% & 0\%\\
        Uncoated / Visible & 25\%~\cite{R5912datasheet} & 2\%\\
        \hline
    \end{tabular}
    \caption{Photon detection efficiencies for the different optical sensors used in our simulations.}\label{tab:PDEs}
    \end{center}
\end{table}

For the VUV light, the pTP-coated X-ARAPUCA time response is modelled after a dedicated Geant4 simulation that accounts for the pTP emission ($\tau_{\text {pTP}}$ = 1.14\,ns) and the inner WLS bar (EJ-286 plastic, $\tau_{\text {EJ-286}}$ = 1.2\,ns) decay times \cite{Marinho:2018doi,Paulucci:2019omb,EljenWLSbar}. For the visible light, in addition to the time response of the TPB described above, the time response of the WLS bar (EJ-280) used inside the uncoated X-ARAPUCAs is modelled with an exponential decay time of 8.5\,ns \cite{EljenWLSbar}, while no bar-response is simulated for the pTP-coated X-ARAPUCA as the EJ-286 bar is transparent to this wavelength. 

\section{Detector response simulation}\label{sec:detresponsesim}

\begin{figure}[t]
\centering
\includegraphics[width=.49\textwidth]{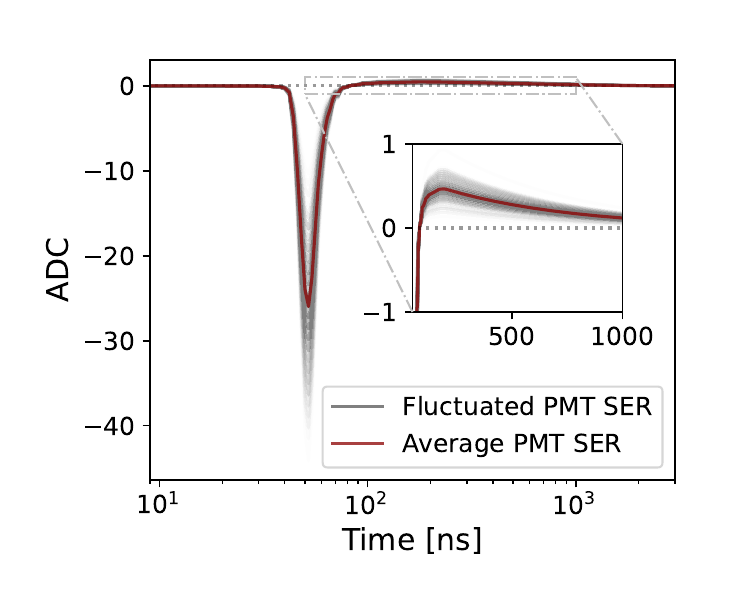}
\includegraphics[width=.49\textwidth]{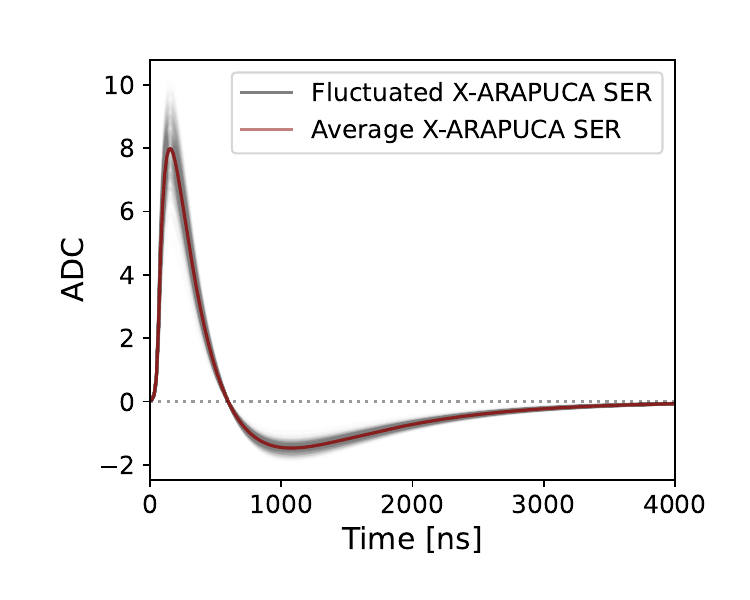}
\caption{Measured single electron response for the PMTs (top) and X-ARAPUCAs (bottom) in SBND. Note the logarithmic scale used for the PMT SER to optimise the simultaneous display of the positive and negative parts of the signal, given the large difference in their duration. The fluctuations included in our simulations, 22\% for the PMTs and 10\% for the X-ARAPUCAs, are also shown.}\label{fig:SERs_SBND}
\end{figure}

The photon detection efficiencies (PDEs) in our simulations are included as global factors on the number of detected photons. Table~\ref{tab:PDEs} summarises the values used in this work, for the different light components and optical detectors. For the PMTs, the deviations from the pure quantum efficiency of the device (25\%) are related to the presence of a WLS in the detection window. For the case of VUV photons, the WLS isotropic emission reduces the detection efficiency by 50\% since visible photons are equally likely to be re-emitted back into the liquid argon volume and away from the PMTs. For the visible photons the transmittance of the TPB has been measured to be 70\%~\cite{SBND_Internal}, reducing the amount of cathode-reflected photons detected by the coated PMTs. Regarding the X-ARAPUCAs, the global efficiency accounts for the WLS coating emission (for the VUV-sensitive modules), dichroic filter transmittance, WLS bar conversion efficiency and SiPM efficiency. While the X-ARAPUCAs have PDEs smaller than PMTs, they offer substantial enhancements over bare SiPMs, making them ideal for instrumenting large areas with tight space constraints.

After amplification and digitisation, each converted photon results in an output signal at the end of our detection chain, known as the single electron response (SER). To make the readout faster and to minimise the number of cables in SBND, the PDs are in AC-coupled configuration, allowing high voltage application and signal readout using a single cable. In this scheme SER signals are bipolar and integrate to zero. Depending on the readout polarity, negative (positive) for our PMTs (X-ARAPUCAs), they show a main pulse followed by an overshoot (undershoot). The SER signals for PMTs and X-ARAPUCAs\footnote{For all the X-ARAPUCAs we use 
a measurement of the Onsemi MICROFC-30050-SMT SiPM~\cite{SiPMs:onsemi} boards with an alternative amplifier and digitiser that produce a similar response to the final one installed in the detector. This does not impact our results as the electronics response is effectively removed during the signal deconvolution as explained in section 5.1.} have been measured at dedicated test-stands. Results are represented in Figure~\ref{fig:SERs_SBND}, where the shaping due to the AC-couplings are clearly visible. For the PMTs (X-ARAPUCAs) the FWHM of the SER is 10\,ns (250\,ns) with an overshoot (undershoot) amplitude of 1.8\% (18.4\%) of the signal peak and extending for about 1.5\,\textmu s (2.9\,\textmu s) before baseline restoring (90-percentile).

Simulated optical waveforms are generated by first scaling the number of photoelectrons (PEs) arriving at each PD by its efficiency from Table~\ref{tab:PDEs}, and then convolving this arrival time distribution with the measured SER described previously. Random fluctuations in the SER integral are applied to better mimic data~\cite{donati2000photodetectors} (see Figure~\ref{fig:SERs_SBND}). For the PMTs, this fluctuation comes from estimations of the gain variation in the first dynode; for the X-ARAPUCAs, the standard deviation of the first PE peak in the spectrum is used. To model electronics noise, we use an uncorrelated additive white Gaussian noise with an RMS of 2.6\,ADC counts for the PMT signals and 0.65\,ADC counts for the X-ARAPUCA signals, corresponding to the measured intrinsic noise of the readout boards.

\begin{figure*}[ht!]
     \centering
     \includegraphics[width=.49\linewidth]{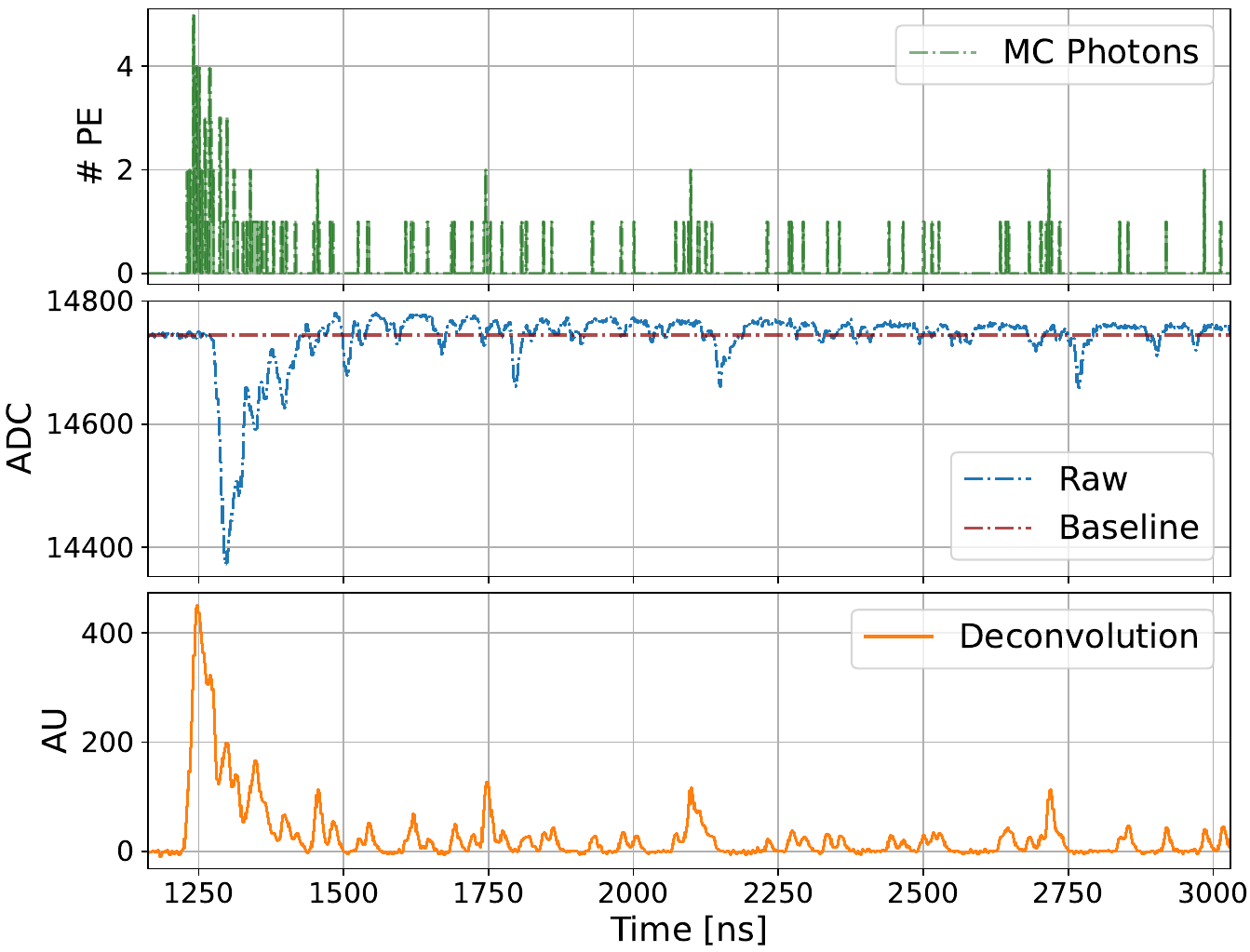} \hfill\insertTextInFigure{-150}{162}{PMT}{8.5}
     \includegraphics[width=.49\linewidth]{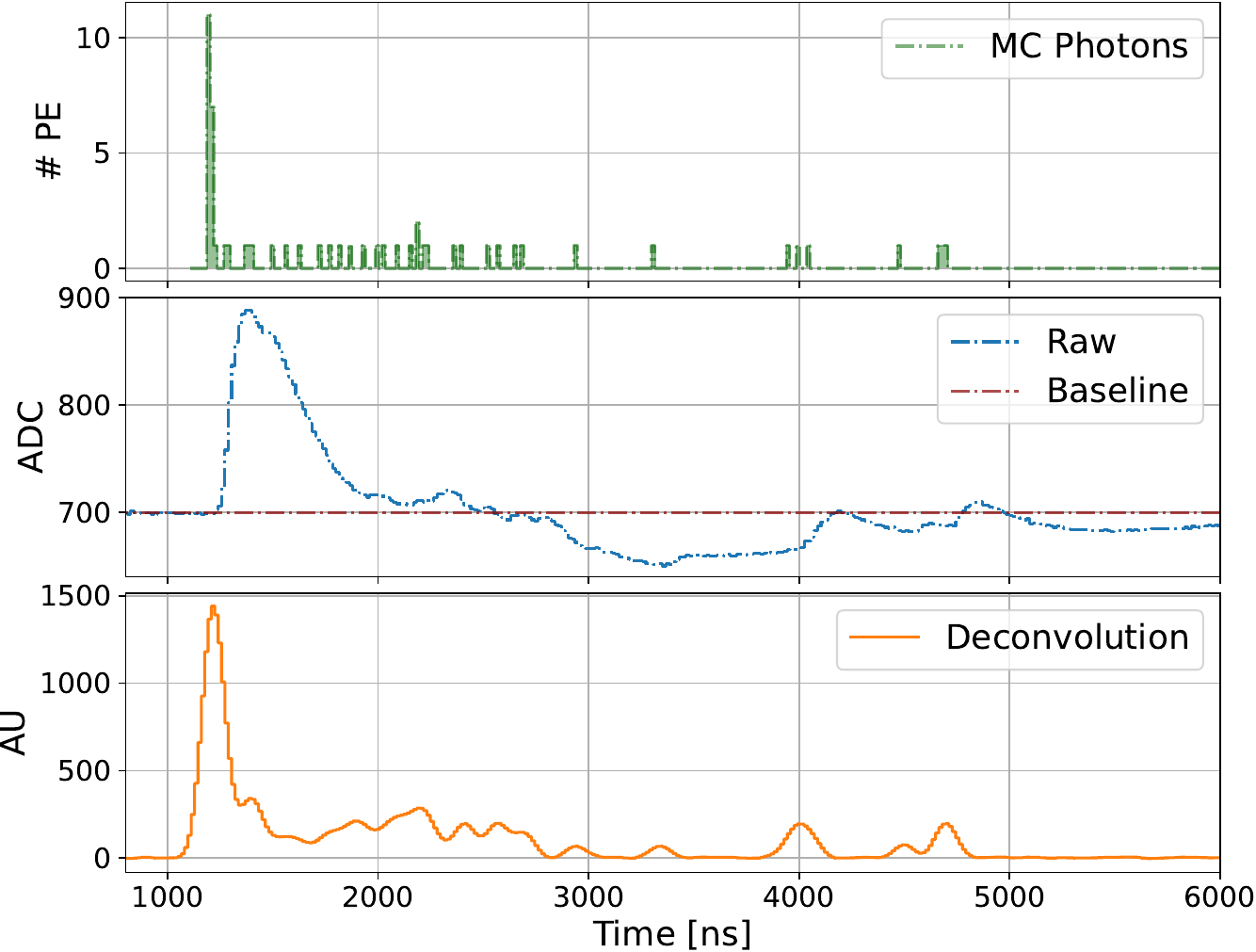}\insertTextInFigure{-150}{162}{X-ARAPUCA}{8.5}
     \caption{Examples of simulated photon arrival time distributions for a neutrino interaction (top), corresponding digitised waveforms (center) and reconstructed unipolar signals after deconvolution (bottom) for a coated PMT (left) and a coated X-ARAPUCA (right).}
     \label{fig:rawsignalecamplePMT}
\end{figure*}

Figure~\ref{fig:rawsignalecamplePMT} shows examples of simulated PMT and X-ARAPUCA waveforms in SBND, before (top) and after (centre) digitisation, where the bipolar shapes in the digitised signals are clearly visible. For the PMTs, we have set the baseline value to minimise possible saturations due to the dynamic range of their 14-bit digitisers, allocating 90\% of the range to the negative peak and 10\% to the overshoot. For the X-ARAPUCAs, we set the baseline of their 12-bit digitisers so that 83\% of the dynamic range is reserved for the positive peak and 17\% for the undershoot. For the PMTs, the observed delay between the photon arrival times (at the photocathode) and the digitised signal (at the anode) accounts for the PMT transit time (55$\pm$2.4\,ns~\cite{R5912datasheet}). In the case of the X-ARAPUCAs the photon propagation time within the module is negligible and the SiPM transit time is at sub-nanosecond level and therefore not simulated.

We also account for the non-linearity of the PMT response at high light intensities. This behaviour has been characterised for the PMT model used in SBND~\cite{Babicz:2019pll,Jetter:2012xp}. To incorporate this effect into our PMT simulation, we employ a data-driven model that effectively accounts for the reduction in the number of PEs reaching each digitisation bin. Finally, we simulate a Dark Count Rate (DCR) of 2000\,Hz~\cite{Belver_2018} in all our PMTs, and 10\,Hz in the X-ARAPUCAs with a 42\% crosstalk~\cite{SBND_Internal}.

\section{Light signal reconstruction}\label{sec:lightreco}

Depending on the detection technology and readout electronics, SER signals can vary significantly. For example, the response to a single photon typically spans around 50\,ns and 500\,ns for the PMTs and X-ARAPUCAs, respectively, as can be seen from the peak time of the waveforms in Figure~\ref{fig:SERs_SBND}. Therefore, requiring different reconstruction algorithms and parameter settings. For this reason, in this work we have studied both systems independently with the goal of showing their performance separately. In future work, we will explore the combination of both systems.

The standard strategy for extracting the number of PEs from a PDS waveform involves a linear area-to-PE conversion. However, as mentioned before, the SBND light signals are bipolar (with a main pulse followed by an overshoot or undershoot) due to the AC-coupled readout. In this situation, multiple photons arriving close in time can shift the waveform baseline, resulting in a large cumulative deviation (see Figure~\ref{fig:rawsignalecamplePMT}-centre panels). The baseline estimation becomes challenging for these bipolar signals, as its value is different from that of the nominal baseline for certain regions of the signal, and the area-to-PE approach fails in estimating the number of photons. Hence a different reconstruction approach is required for an accurate PDS signal processing in SBND. Next we describe a deconvolution-based method which aims to remove the signal bipolarity, providing an estimator for the number of PEs and their arrival times.

\subsection{Waveform deconvolution}\label{subsec:DecoAlg}

If a completely linear response is assumed for the PDs, the raw signals $f(t)$ are the convolution of the photon arrival time distribution $s(t)$ and the detector response $r(t)$:
\begin{equation}
f(t) = s(t) \ast r(t) \equiv\int_{-\infty}^{\infty}s(t)r(t-\tau)d\tau.
\label{eq:convolution}
\end{equation}
This makes possible the recovery of our true signals by using the convolution theorem:
\begin{equation}
s(t)=\mathcal{F}^{-1}\left\{\frac{F(\omega)}{R(\omega)}\right\}\label{eq:decoformula},
\end{equation}
where $F(\omega)$ and $R(\omega)$ represent the Fourier transforms $\mathcal{F}$ of $f(t)$ and $r(t)$, respectively. However, this simple approach starts to degrade when we take into account the presence of noise. Therefore, for the correct processing of our signals we implement different strategies to mitigate noise. For the finely-sampled PMT signals, 
we apply a two-step waveform smoothing before the Fourier transform~\cite{aguilararevalo2021dark}: 
\begin{enumerate}
\item Exponential average smoothing:\newline $\tilde{f}_i=(1-\alpha)\tilde{f}_{i-1}+\alpha f_i$
\item Unweighted average smoothing:\newline $\tilde{\tilde{f}}_i=(\tilde{f}_{i-1}+\tilde{f}_{i}+\tilde{f}_{i+1})/3$,
\end{enumerate}
where $f_i$ stands for the waveform value at a given time-tick $i$, and $\tilde{f}_i$ and $\tilde{\tilde{f}}_i$ stand for the smoothed value after steps 1 and 2 respectively; with $\alpha$ a free parameter set to 0.3 after an empirical optimisation. For the X-ARAPUCA system the smoothing is not needed as the slower sampling already works as a low-pass filter. Next, for both systems we apply a filter in the frequency domain ($G(\omega)$) to maximise our signal-to-noise ratio, leaving Equation~\ref{eq:decoformula} in the following form:
\begin{equation}
s(t)=\mathcal{F}^{-1}\left\{G(\omega)\frac{F(\omega)}{R(\omega)}\right\}\label{eq:decoformula2}.
\end{equation}
Among the different options available we have chosen a Gaussian filter due to its simplicity and good performance (see section~\ref{subsec:opflashreccomp})
\begin{equation}\label{eq:gauss-filter}
    G(\omega)=e^{-\frac{1}{2}(\frac{\omega}{\omega_c})^2}\ (\omega>0),
\end{equation}
with $\omega_c$ a free parameter that is set by fitting Equation~\ref{eq:gauss-filter} to the Wiener filter~\cite{wienerbook} estimated for the SER signals. The best fit result obtained is $\omega_c=49.0\pm0.2$\,MHz for the PMTs and $\omega_c=3.2\pm0.3$\,MHz for the X-ARAPUCAs. Figure~\ref{fig:rawsignalecamplePMT}-bottom shows one example for a deconvolved signal, following the procedure described above, for both a PMT and an X-ARAPUCA, where it can be seen how the deconvolution has eliminated the bipolarity of the signals while maintaining the relative size and temporal position of the peaks.

\subsection{Optical hit and flash objects}\label{subsec:OpHit-flash}

\begin{figure}[t]
    \centering
    \includegraphics[width=1.\linewidth,  angle=0]{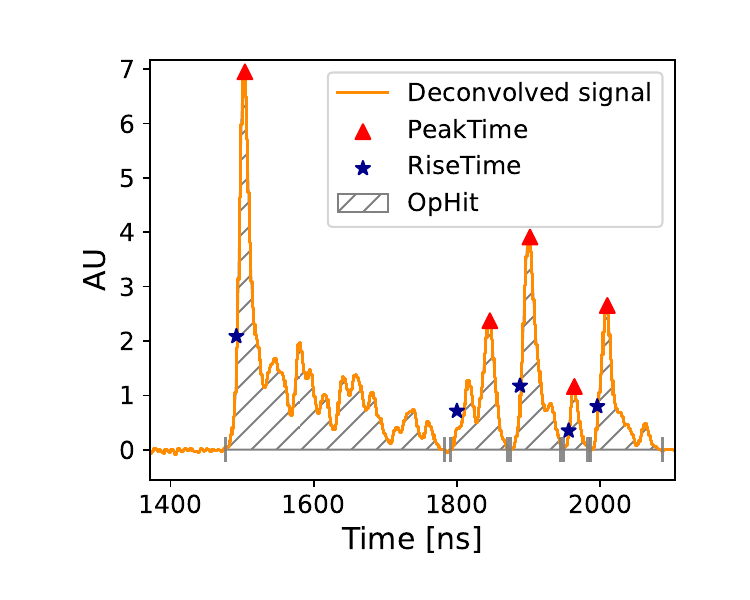}
     \caption{Example of the OpHit finder algorithm performance for a PMT deconvolved waveform. The start/end point of the OpHits are indicated by the vertical lines along the baseline. The different estimates available for the OpHit-times are also shown.}
     \label{fig:ophitfinder_algo}
\end{figure}

The next step in the optical reconstruction is to recover each scintillation photon arriving to the optical channels by looking for pulses along the deconvolved waveforms. These optical hits (OpHits) are the basic objects in the optical reconstruction. Following the subtraction of the baseline, that we estimate using the start and end portion of the deconvolved signals (400\,ns on each side), pulses are found by identifying samples with a value higher than 1/4 the amplitude of the deconvolved SER signal and 3 times the standard deviation of the baseline RMS. Figure~\ref{fig:ophitfinder_algo} illustrates the performance of the OpHit finder algorithm for a PMT waveform. It can be seen how multiple peaks are merged into a single OpHit when multiple photons arrive simultaneously (or very close) to the PD. This is particularly relevant for the fast light.

\begin{figure*}[t!]
    \centering
     \includegraphics[width=1.\textwidth]{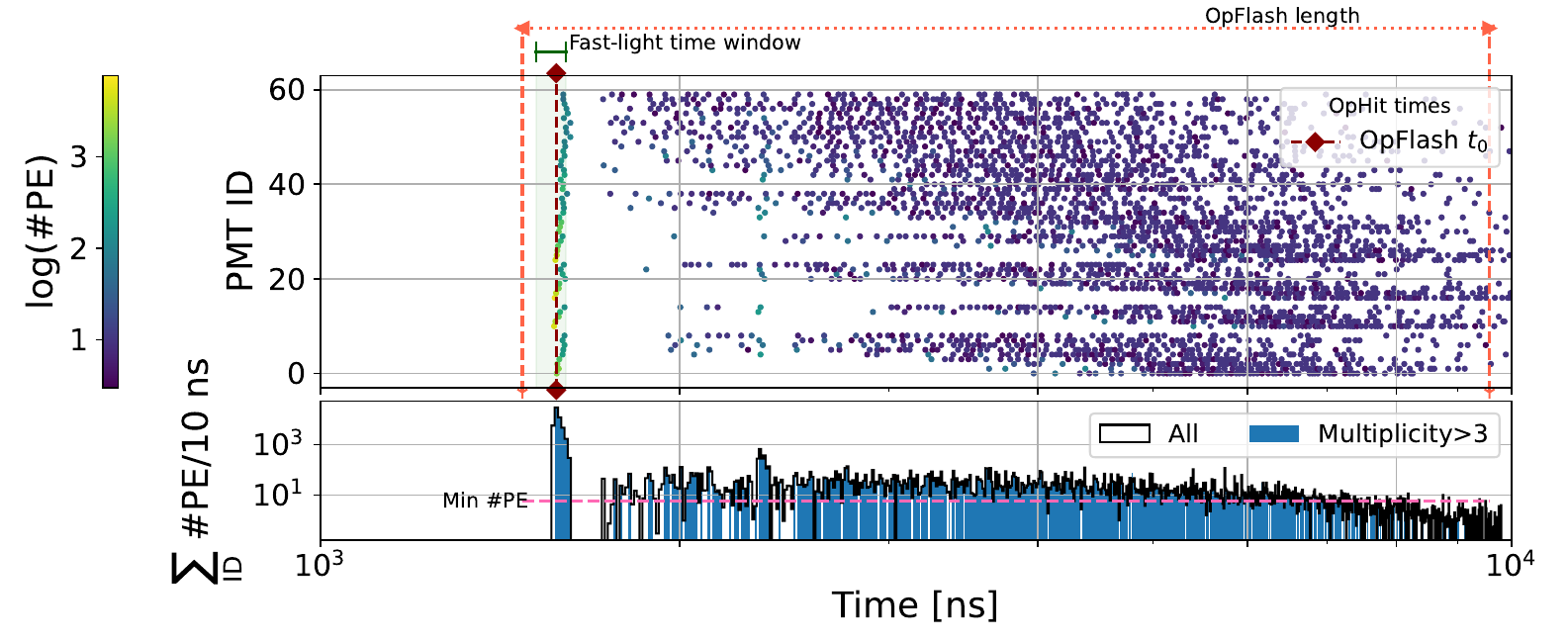}
     \caption{
     Illustration of the OpFlash finder reconstruction algorithm. The dotted orange (solid green) horizontal line represents the OpFlash length (fast emission) window parameter. The gap observed after the fast emission is caused by the large width of the OpHits reconstructed for this light component. The solid histogram (OpFlash X-axis projection) represents the time intervals satisfying the multiplicity condition along the OpFlash length. The dashed pink line shows the minimum number of PEs per optical channel required to identify an OpFlash.} \label{fig:flashfinder_algo}
\end{figure*}

The number of PEs in the signals is recovered by the integral of the OpHits. The possible broadening of the OpHits due to the simultaneous arrival of photons, together with the smearing resulting from the deconvolution process, can introduce biases in the estimation of the arrival time of photons from these objects. To minimise these issues, in the PMT system we define the OpHit-time as the time sample in which the deconvolved signal goes above a certain fraction of the maximum peak value. Using a rise-time threshold of 15\% we obtain a resolution of about 1.6\,ns in the estimation of the arrival time of the first photon contributing to each OpHit. For the X-ARAPUCA system, due to the coarser sampling, a Gaussian function is fit to the deconvolved waveform to estimate its peak time, achieving a resolution of 5.9\,ns.

We use the sum of the OpHits to reconstruct the amount of light generated by a given interaction within each TPC. We will refer to clusters of OpHits as optical flashes (OpFlashes). An OpFlash is built if a multiplicity condition is satisfied: a minimum number of 6\,PE is detected by more than 3 optical channels within a time interval of 10\,ns for the PMTs or 60\,ns for the X-ARAPUCAs. The length of an OpFlash is set to 8\,\textmu s to account for the total amount of light generated in the interaction (see Figure~\ref{fig:scintemission}-Right).

\begin{table}[t!]
\begin{center}
\begin{tabular}{ |c|c| } 
 \hline
\textbf{Parameter} & \textbf{Value} \\
 \hline
 Minimum number of PEs & 6\,PE \\
 Minimum number of optical channels & 3 ch\\  
 Time interval (PMT / X-ARAPUCA) & 10\,ns / 60\,ns \\
 PEs threshold & 20\,PE \\
 OpFlash length & 8\,\textmu s\\
 Time window for t$_0$ & 30\,ns \\
 Veto-window & 8\,\textmu s\\
 \hline
\end{tabular}
\end{center}
\caption{List of OpFlash parameters used in the SBND optical reconstruction. All the parameters are shared by PMTs and X-ARAPUCAs, except the time interval due to the different behaviour of the two systems.} \label{tab:OpFlash}
\end{table}

The start time of an OpFlash represents the time of the interaction (t$_0$), so it must be reconstructed with great care. To obtain the value of t$_0$ we average the OpHit times for the channels that collectively contribute 50\% of the total light within a 30\,ns window around the time interval with the largest number of PEs. Once an OpFlash candidate has been defined, it will only be saved if the integrated number of reconstructed PEs is above 
20\,PE. This cut is set to avoid selecting candidates originating from dark counts or low energy backgrounds like $^{39}$Ar. Finally, a 8\,\textmu s veto-window is also applied to prevent other flashes being created during the OpFlash duration, although multiple flashes can be found within an event. Figure~\ref{fig:flashfinder_algo} illustrates the OpFlash reconstruction procedure. Table~\ref{tab:OpFlash} summarises the parameters used to build OpFlash objects in SBND. The value of the parameters used in the reconstruction of both OpHits and OpFlashes are the result of an optimisation process which seeks to maximise the amount of useful information from the light signals and the number of reconstructed neutrino events, but their final values may be revisited once data becomes available.

\section{Reconstruction efficiency}\label{sec:reco_effficiency}

\begin{figure*}[ht!]
    \centering
     \includegraphics[width=.49\textwidth]{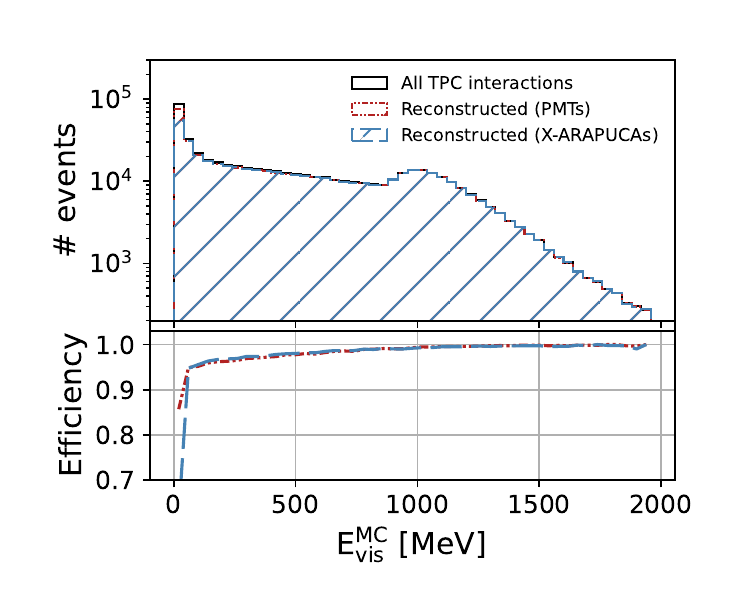} 
     \includegraphics[width=.49\linewidth]{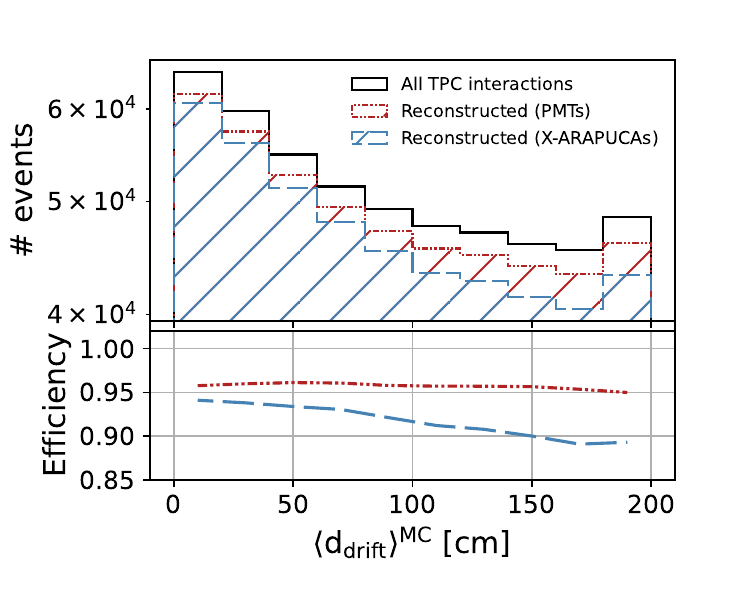}
     \caption{OpFlash reconstruction efficiency for a sample of BNB neutrino + cosmic events as a function of the deposited energy (left) and drift distance (right) for PMTs (red) and X-ARAPUCAs (blue). The dashed (solid) lines represent the number of events with a reconstructed OpFlash (all the events). The shape of the distributions on the left is the combination of the BNB spectrum, cosmic spectrum and detector acceptance, with the bump around 1000\,MeV caused by muons crossing the detector volume.} \label{fig:flashrecoeff}
\end{figure*}

In this section we show the overall efficiency of reconstructing the light signals in SBND. 
To carry out these studies we have used a sample of 30,000 BNB neutrino events simulated in LArSoft and using GENIE~\cite{Andreopoulos:2009rq,Andreopoulos:2015wxa} for the generation of neutrino interactions. We have excluded waveforms where saturation occurs. On average, for the PDEs considered in this work, this translates into not using 2 PMT and 1 X-ARAPUCA waveforms per BNB event, and therefore we expect the impact to be negligible.

\subsection{OpFlash reconstruction efficiency} \label{subsec:opflashreceff}

An efficient reconstruction of the OpFlash objects is of great importance as they represent the light associated with an interaction. To study their reconstruction efficiency, we have used our neutrino sample together with the corresponding cosmic ray overlay, since we are interested in studying the reconstruction efficiency of neutrino events in the presence of the cosmic background, as will be the case for real data. We have simulated the cosmic ray interactions in LArSoft using Corsika~\cite{Heck:1998vt} as the event generator. In this study we have also considered an interaction as visible if its energy deposition is larger than 5\,MeV, excluding the events at lower energies.

The OpFlash reconstruction efficiency, defined as the ratio between the number of interactions with a reconstructed OpFlash and the total number of interactions, is shown in Figure~\ref{fig:flashrecoeff} as a function of the deposited energy and average drift distance ($\rm d_\text{drift}$)\footnote{Hereafter, with the label MC we will refer to variables that have been obtained using true level information.}. 
As expected, the efficiency drops at low energy depositions. However, some flashes are also lost at high energy values. They correspond to {\it in-time} interactions (neutrino-cosmic or cosmic-cosmic) occurring within the OpFlash time length. As explained in section \ref{subsec:OpHit-flash}, a veto window is applied after identifying an OpFlash, meaning that no other OpFlash can be claimed during the veto time. In these cases only one OpFlash is recovered, typically initiated by the interaction producing the larger amount of photons, but with photons coming from the two interactions (OpFlash pile-up). The reduction of efficiency due to coincident events has been estimated to be around 2\% for both PMTs and X-ARAPUCAs. The dependence of the reconstruction efficiency with the drift distance is very small (within 1\% for the PMTs and 5\% for the X-ARAPUCAs). Overall, we obtain a global OpFlash reconstruction efficiency of 95.8\% with the PMT system and 92.2\% with the X-ARAPUCA system.

\subsection{OpFlash reconstruction completeness} \label{subsec:opflashreccomp}

\begin{figure*}[t]
    \centering
    \includegraphics[width=.49\textwidth]{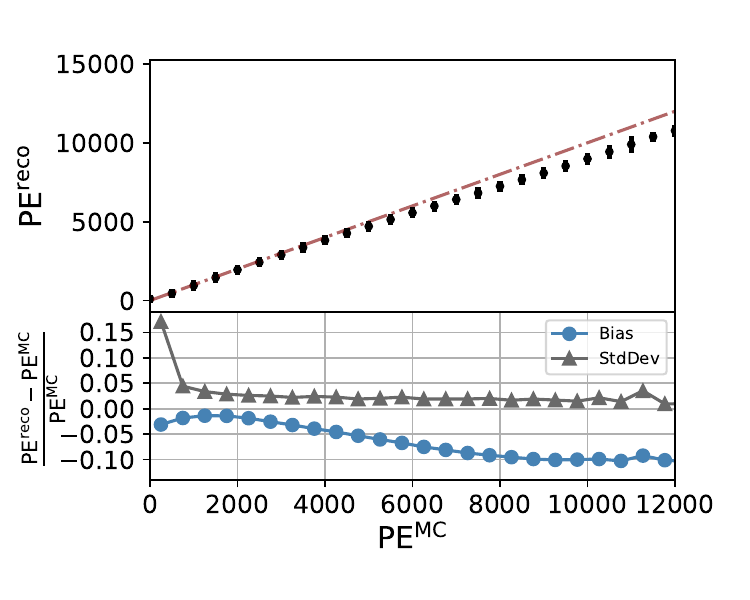}\insertTextInFigure{-160}{142}{PMT}{8.5}
    \includegraphics[width=.49\textwidth]{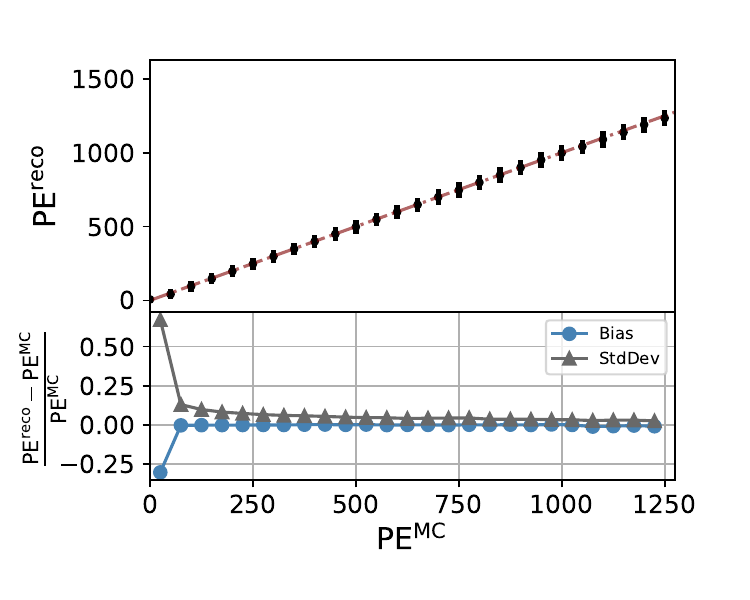}\insertTextInFigure{-160}{142}{X-ARAPUCA}{8.5}
    \caption{Number of reconstructed PEs from OpHits within an OpFlash, and accuracy ($\equiv$ Bias) and resolution ($\equiv$ StdDev), as a function of the total number of simulated PEs within one channel in the PMT (left) and X-ARAPUCA (right) systems.
    The non-linear behaviour of PMTs is clearly visible from 3000\,PEs onwards.}\label{fig:pecompleteness}
\end{figure*}

To study the resolution in the reconstruction of the number of PEs, we use the OpHit objects within an OpFlash. We compare the integral of the OpHits at each PD ($\rm PE^{reco}$) with the total number of simulated photons arriving to the same channel ($\rm PE^{MC}$). Results are shown in Figure~\ref{fig:pecompleteness}. In the PMT system we obtain an almost flat resolution better than 3\% for channels with more than 1000\,PEs. The trend of the bias shows the non-linearity of the PMTs starting at about 3000\,PEs, where we go from underestimations smaller than 2\% to values between 8-10\% for channels with more than 6000\,PEs. For the X-ARAPUCAs, the number of photons is reduced as their area and efficiency are smaller than those of PMTs, achieving a resolution and bias better than  6\% and 1\% respectively, for channels with more than 250\,PEs.

\section{Reconstruction performance}\label{sec:reco_performance}

After assessing the reconstruction efficiency of our primary reconstructed objects (OpFlashes) and the resolution in the reconstruction of the number of PEs, in this section we proceed to evaluate the detector performance based on several key metrics derived from the OpFlash objects. 
\subsection{Light yield}\label{subsec:light_yield}

\begin{figure*}[t]
    \centering
    \includegraphics[width=.49\textwidth]{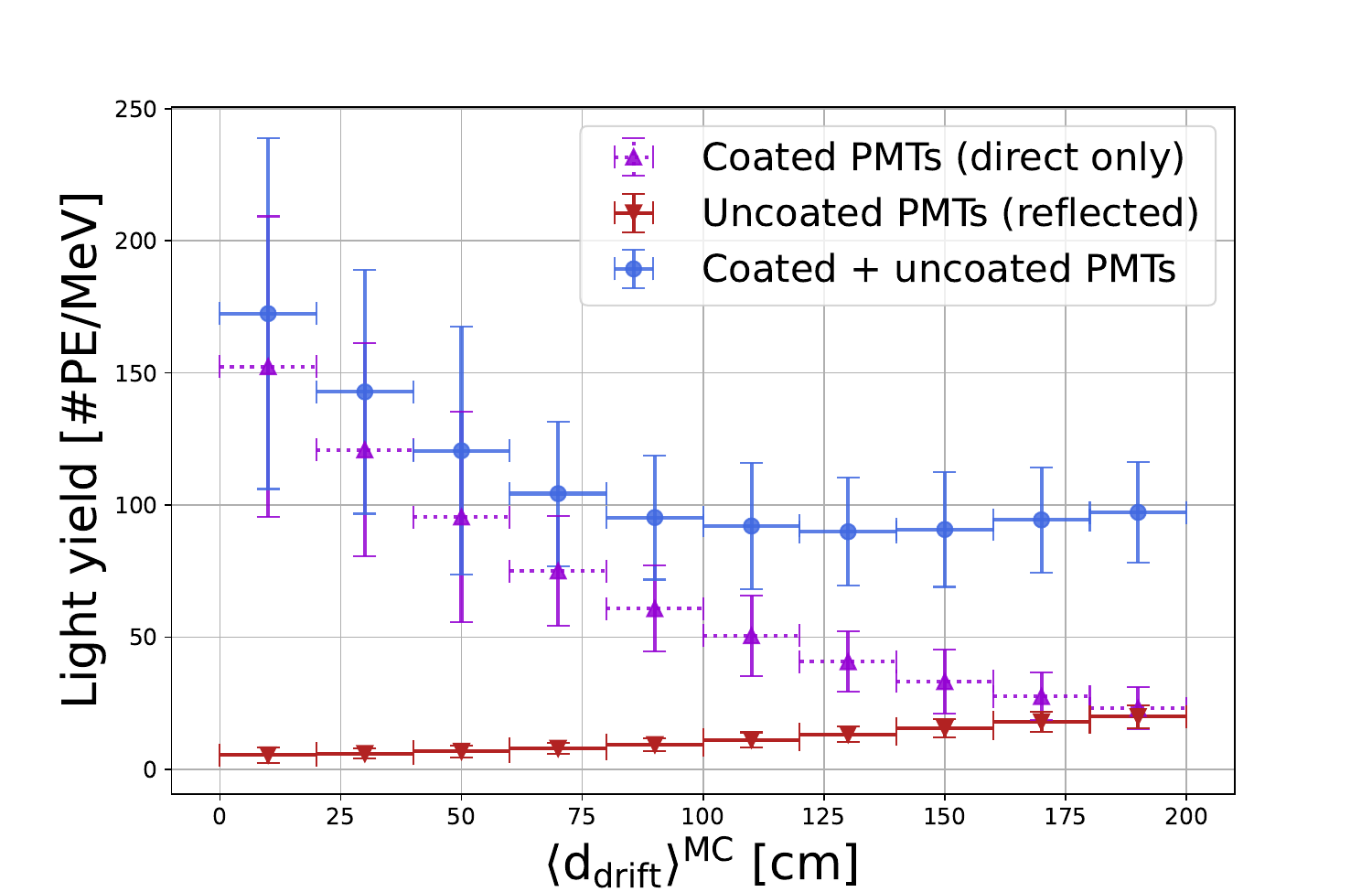}
    \includegraphics[width=.49\textwidth]{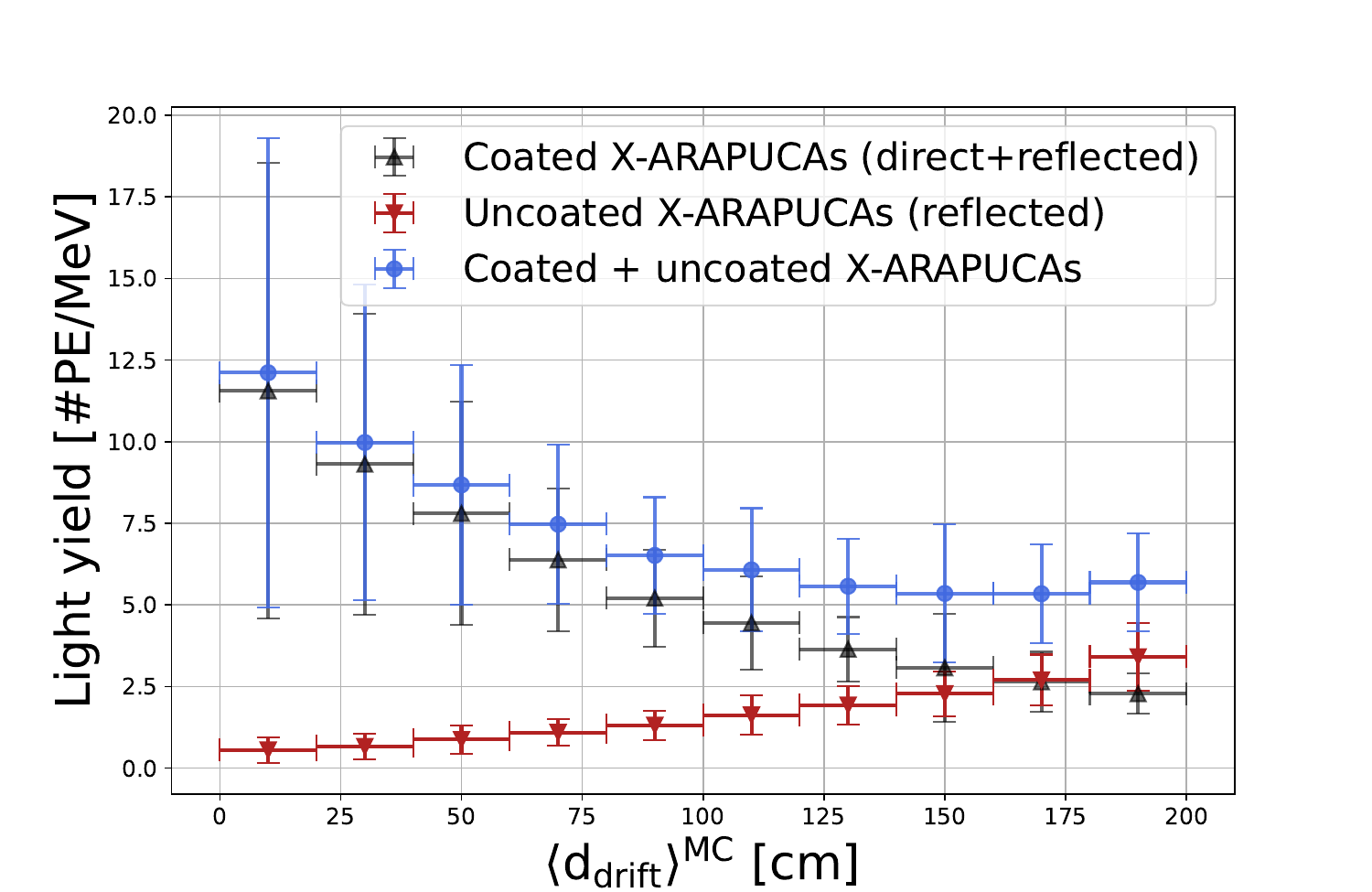}
    \caption{Expected LY in SBND as a function of the mean drift distance for the different PMT (left) and X-ARAPUCA (right) flavours, according to the detection efficiencies in Table~\ref{tab:PDEs}. Note that the large size of the error bars as we approach the detection plane is primarily due to border effects (large difference in the registered light between events occurring near or far from the edges of the detector) and not to uncertainties in the simulation. }\label{fig:LYreco}
\end{figure*}

The expected LY for each optical detector type is shown in Figure~\ref{fig:LYreco} as a function of the drift distance. The number of reconstructed PEs is obtained from the OpFlash objects, while the amount of deposited energy and its average drift position are taken from truth-level information. The total LY from uncoated PMTs is much lower than that of coated PMTs, not only because they are only sensitive to the reflected light component, but also because they are in a ratio 1:4. For the X-ARAPUCAs, since there are equal number of coated and uncoated units, this asymmetry is not present and the difference is simply due to the larger total amount of light detected by the coated ones. When comparing the two systems, the PMTs collect more light as their PDEs are higher and have a larger photo-coverage than the X-ARAPUCAs (9.6\% and 7\% of the anode plane, respectively). As no PMT is VUV-only sensitive, the LY for the direct light in Figure~\ref{fig:LYreco}-Left (points with dashed error bars) has been simply estimated by subtracting to each coated signal its closest uncoated one (after correcting for the different efficiencies)\footnote{At each PDS box (see Figure~\ref{fig:PDS}) the distance between the central (uncoated) PMT and the corner (coated) PMTs is 50\,cm.}. This gives an estimate of the PMT system without the TPB-coated reflector foils on the cathode.

As expected, the LY for the uncoated PDs increases as the interactions approach the cathode (d$_\text{drift}$= 200\,cm), where visible light is re-emitted from the WLS reflectors. On the other hand, the closer to the anode, the more direct light is collected. The anti-correlation between the two components makes the LY significantly larger and more uniform than using the direct component alone, which is a highly desirable behaviour that SBND is able to achieve thanks to the innovative design of its PDS. In particular, the fraction of light gained by the coated PMTs ranges from 50\% in the centre of the TPC to 400\% in the region close to the cathode. For the coated X-ARAPUCAs the gain is much more modest as the choice of the dichroic filter and WLS bar is not well suited for trapping visible photons, but it is compensated by the fact that the other half of the system is dedicated to these wavelengths. In the X-ARAPUCAs, the improved uniformity comes from the sum of both coated and uncoated units (blue-circle points in Figure~\ref{fig:LYreco}-Right). The large error bars\footnote{All error bars in this work represent the standard deviation of the distribution of points in each case.} in Figure~\ref{fig:LYreco} are driven by border effects, as the total light collected can vary significantly for energy depositions near or far from the edges of the active volume, even if they are at the same drift distance.

\subsection{Position resolution}\label{sec:position}

\begin{figure*}[t]
    \centering
         \includegraphics[width=.5\textwidth]{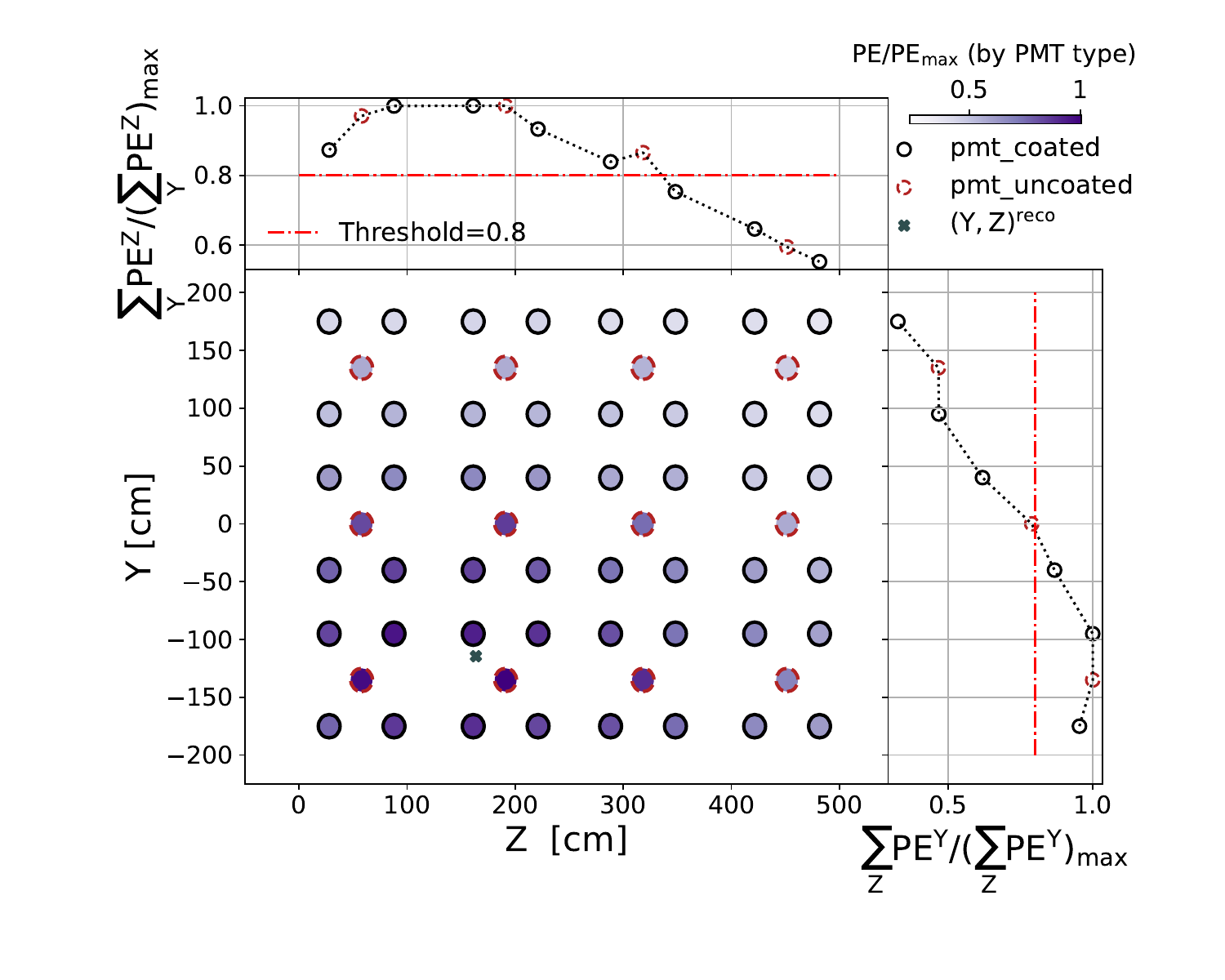}
         \includegraphics[width=.47\linewidth]{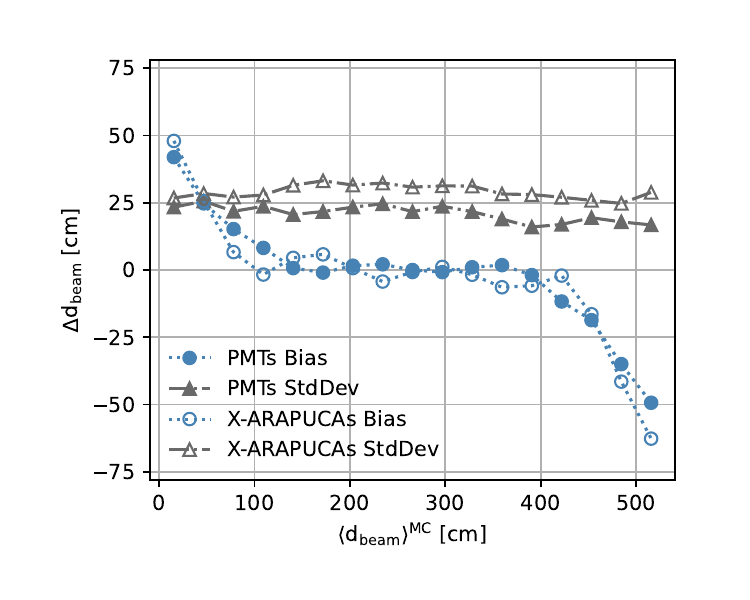}
     \caption{Left: Illustration of the (Y, Z) reconstruction procedure. Right: Accuracy and resolution in the estimation of the interaction point in the beam direction (Z-coordinate) using only scintillation light for PMT (solid markers) and X-ARAPUCA (empty markers) flashes. The difference in resolution is mainly caused by the higher LY of the PMT system.} \label{fig:Z-reco}
\end{figure*}

An important application of the scintillation signals in a liquid argon neutrino detector is the determination of the position of the events that generate the light flashes. The location of events using only the PDS will never be as good as that obtained using the TPC information (which needs an external time reference to resolve the degeneracy along the drift coordinate). Even so, an estimation of the position with light is extremely important in detectors located near the surface as it allows us to match ionisation patterns to the precisely timed light signals, which will help identify which interactions happened inside of the beam gate (signal candidate) and which happened outside of it (background). 

In standard LArTPCs, the light-based position reconstruction is generally performed in the coordinates defining the PDS plane (Y-Z plane in our coordinate system). The achieved resolution will then be largely determined by the spacing between adjacent optical channels. 

Taking advantage of the high PD-density in SBND we have developed a simple threshold algorithm to reconstruct the (Y, Z) of the interaction using only light signals. We perform the PE summation at constant Y and Z positions for each PD type. For the position estimation, we only consider the PDs whose PE summation at each Y/Z position deviates less than a certain fraction from the one with the most light. By an optimisation process, using a sample of BNB-like neutrinos, we set a value of 20\% for this difference. Figure~\ref{fig:Z-reco}-Left illustrates the (Y, Z) reconstruction procedure for a simulated neutrino event using the PMT system. Each circle represents a PMT in the PDS-plane with a colour gradient showing the signal integral relative to the maximum. The fixed threshold at 80\% of the maximum is also shown by dashed lines. In Figure~\ref{fig:Z-reco}-Right we show the resolution we obtained for the reconstruction of the interaction point along the beam direction d$_{\text{beam}}$ (Z) with both systems. We observe minimal bias and a resolution of 25\,cm within 1\,m inside the detector volume. Outside that region, some border effects appear that bias our estimate up to a value of 50\,cm for interactions occurring just at the boundary of the active volume. Similar results are obtained for the Y position reconstruction.

\begin{figure}
    \includegraphics[width=0.49\textwidth]{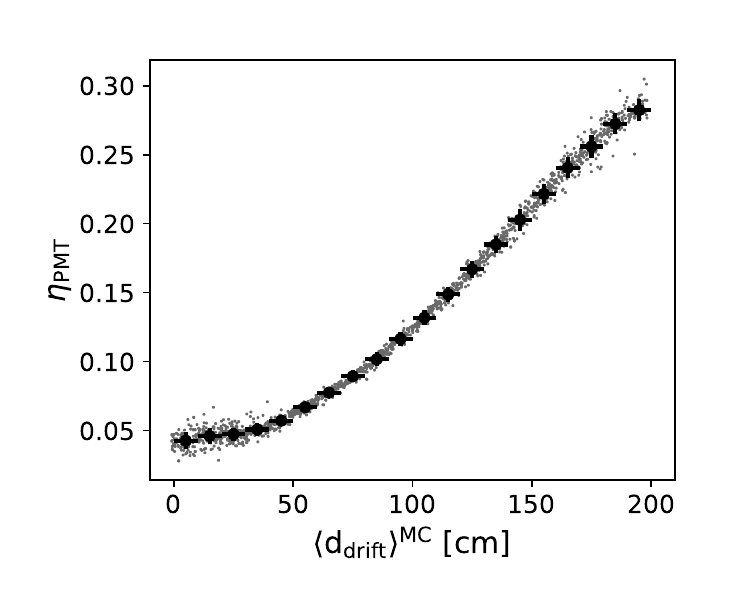}
    \includegraphics[width=0.49\textwidth]{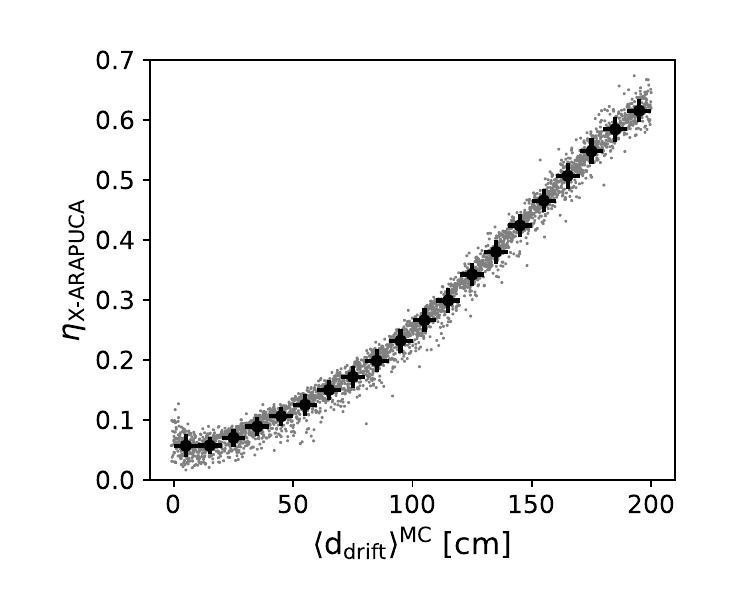} 
    \includegraphics[width=0.49\textwidth]{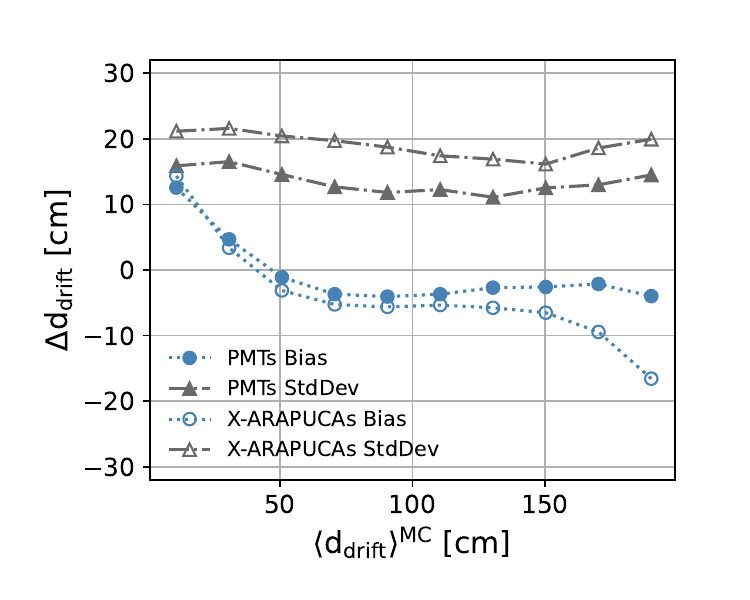} 
    \caption{Top and middle: Calibration curves for the $\eta_{\rm _{PMT}}$ and $\eta_{\rm _\text{X-ARAPUCA}}$ parameters. Bottom: Accuracy and resolution in the estimation of the drift distance using the $\eta_{\rm _{PMT}}$ (solid markers) and $\eta_{\rm _\text{X-ARAPUCA}}$ (empty markers) parameters.}
    \label{fig:fX-calib}
\end{figure}

A unique feature of SBND's PDS design is its ability to reconstruct the mean drift distance at which the energy is deposited. This can be done by exploiting the correlation between $\langle$d$_\text{drift} \rangle$ and the relative amount of photons of the direct and reflected light components measured for a given interaction. For our PMT system, we define the $\eta$ parameter as the ratio between the number of photons seen by the uncoated and the coated PMTs,
\begin{equation}
    \eta_{\rm _{PMT}}\equiv\frac{\sum \rm PE_{ uncoated}}{\sum \rm PE_{ coated}}.
   \label{eq:eta-pmt}
\end{equation}
For the X-ARAPUCA system, given the low efficiency for visible photons of their coated devices,
to reduce statistical fluctuations in the denominator for events near the cathode, we add $\rm \sum PE_{uncoated}$ so the $\eta$ parameter is defined as
\begin{equation}
    \eta_{\rm _\text{X-ARAPUCA}}\equiv\frac{\sum \rm PE_{ uncoated}}{\sum \rm PE_{ coated}+\sum \rm PE_{ uncoated}}.
   \label{eq:eta-xa}
\end{equation}

Both quantities can be directly obtained from the OpFlash reconstructed number of PEs. Figures~\ref{fig:fX-calib}-Top and Middle show the correlation between $\langle$d$_\text{drift} \rangle$ and the $\eta$ parameters. These calibration curves were obtained from a simulated sample of cosmic muons with a well defined drift coordinate, {\it i.e.} muon tracks contained in narrow (10\,cm) slices along the drift. In practice, these tracks can be selected in SBND data using the external CRT system that can directly trigger on these muon topologies with a resolution better than 2\,cm. Figure~\ref{fig:fX-calib}-Bottom shows the resolution obtained in the estimation of the drift distance for our sample of BNB-like neutrinos using the $\eta$ parameters. For the PMT system we see a small bias ($<$\,5\,cm) and a resolution better than 15\,cm for distances larger than 50\,cm. The loss of sensitivity at the shorter distances is mainly driven by two effects: the PMT non-linearity for large signals, and energy depositions happening inside and outside the active volume where the probability of detecting photons generated on those regions is quite different. Again, the slightly worse result of the X-ARAPUCAs is mainly due to the lower LY of this system.

\subsection{Timing resolution}\label{sec:timeres}

In a LArTPC, light travels about a million times faster than charge, so it is the signal of choice to indicate the time of the interaction. To quantify the timing resolution of our system, we define $\rm \Delta t_0$ as the difference between the OpFlash time and the true interaction time. A constant and homogeneous timing resolution throughout the detector requires accounting for the light propagation time from the energy depositions to the PDs. For this purpose, and bearing in mind that visible light travels almost twice as fast as VUV light in LAr, we will distinguish among three different regions in our detector:
\begin{figure}[t]
    \centering
    \includegraphics[width=1.\linewidth, angle=0]{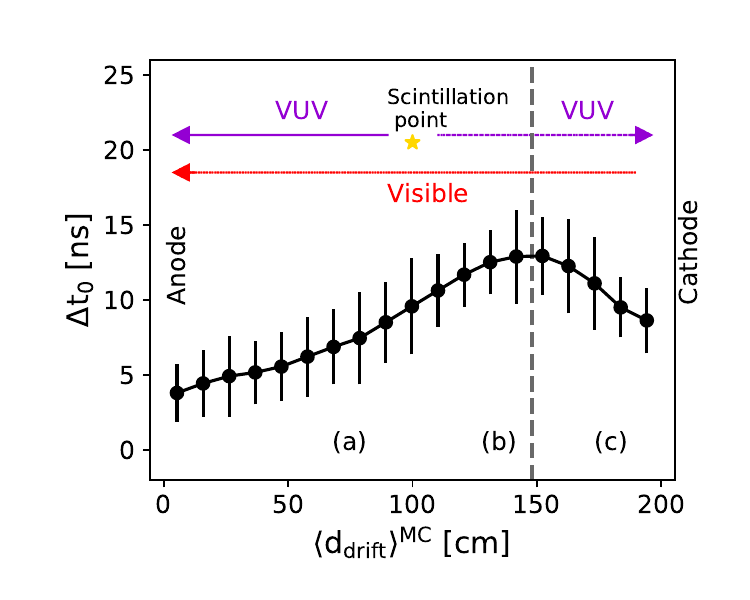}
    \caption{OpFlash time difference to the interaction time, as a function of the drift distance using the PMT system. The error bars correspond to the standard deviation in each drift bin. A cartoon of the two light component paths is also shown: VUV photons can either propagate from the energy deposition point (yellow star) directly to the PDS or propagate to the cathode where they are re-emitted with visible wavelengths. The tipping point at about 45\,cm from the cathode is clearly visible.}
     \label{fig:lightprop_cartoon}
\end{figure}

\begin{description}
    \item{\bf (a)} Direct-component dominated region, where the first photons arrive from the direct component (VUV wavelengths). The time of the photons increases linearly with the distance to the anode.
    \item{\bf (b)} Tipping point, or drift distance where the photons from the two light components arrive at the same time to the PDs. At this point the propagation time reaches its maximum value.
    \item{\bf (c)} Reflected-component dominated region, where the first photons arriving to the PDs come from the re-emitted component (visible wavelengths). In this region, the time of the photons increases with the distance to the cathode (light has to travel as VUV photons from the interaction point to the cathode, where they are wavelength shifted, and then propagate back along the 200\,cm of maximum drift distance with visible wavelengths).
\end{description}
An illustration of the arrival path of the two light components to the PDS can be seen in Figure~\ref{fig:lightprop_cartoon}. The points represent the values of $\rm \Delta t_0$ from our PMT system and for our neutrino sample as a function of the drift distance. It can clearly be seen that $\rm \Delta t_0$ follows the trends described above, ranging from a few to about 15\,ns.

\begin{figure*}[ht!]
    \centering
         \includegraphics[width=.49\textwidth]{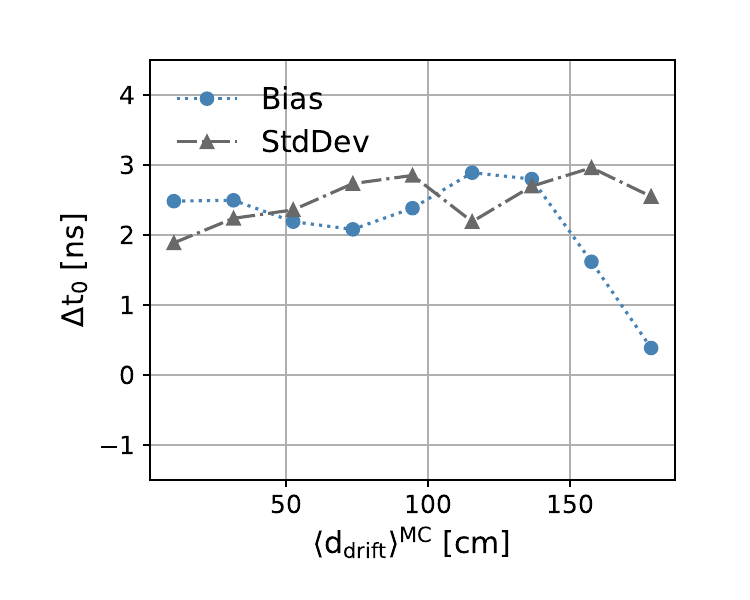}
         \includegraphics[width=.49\linewidth]{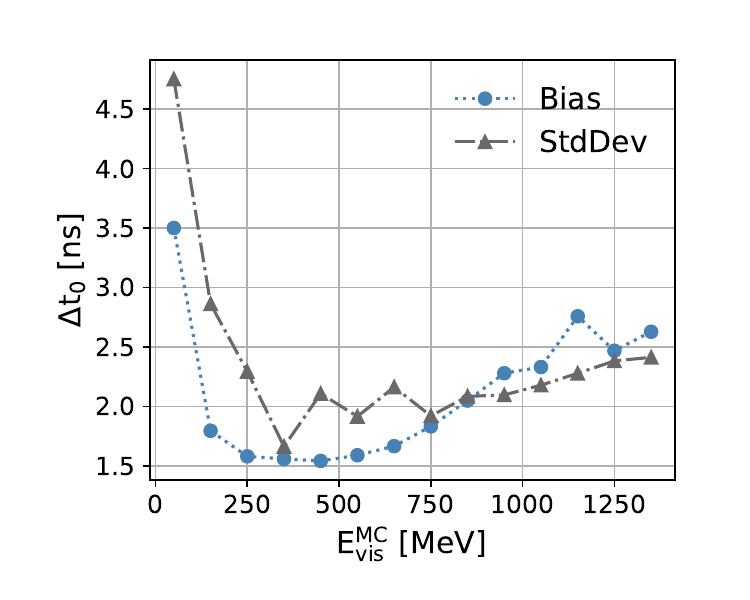}  
     \caption{Time accuracy and resolution of the PMT system as a function of the drift distance (left) and deposited energy (right), after corrections for propagation effects.} \label{fig:ResoTime}
\end{figure*}

The average drift coordinate of the interaction $\langle \text{X$_\text{reco}$} \rangle$ can be estimated from the calibration curves in Figure~\ref{fig:fX-calib} taking as input the measured $\eta$ value. Once $\langle \text{X$_\text{reco}$} \rangle$ has been estimated, and defining the tipping point as
\begin{equation}
\text{X}_\text{T}\equiv\frac{\text{X}_\text{PDS}}{2}\left(1-\frac{\text{V}^\text{VUV}_\text{group}}{\text{V}^\text{Vis}_\text{group}}\right),
\end{equation}
we can correct for the light propagation delay as follows:
\begin{equation}
    \text{T}_\text{OpFlash} \rightarrow
    \begin{dcases}
        & \text{T}_\text{OpFlash} -\frac{\text{X}_\text{PDS}-\langle \text{X$_\text{reco}$} \rangle}{\text{V}^\text{VUV}_\text{group}}\\
        &\text{\quad if } \langle \text{X$_\text{reco}$} \rangle > \text{X}_\text{T},\\
        & \text{T}_\text{OpFlash} - \left(\frac{\langle\text{X}_\text{reco}\rangle}{\text{V}^\text{VUV}_\text{group}}+\frac{\text{X}_\text{PDS}}{\text{V}^\text{Vis}_\text{group}}\right)\\ 
        &\text{\quad if }  \langle \text{X}_\text{reco}\rangle<\text{X}_\text{T},
    \end{dcases}
\end{equation}
where $\text{V}^\text{VUV}_\text{group}$ and $\text{V}^\text{Vis}_\text{group}$ are the group velocity for the VUV and visible photons\footnote{We have used $\text{V}^\text{VUV}_\text{group}$=13.5\,cm/ns and $\text{V}^\text{Vis}_\text{group}$=23.9\,cm/ns.
}, respectively, and $\text{X}_\text{PDS}$ is the location of the PDS in the drift direction.

After correcting for the photons time-of-flight (ToF) delay, the final $\rm \Delta t_0$ resolution (both bias and standard deviation) obtained in SBND for the PMT system using only scintillation light can be seen in Figure~\ref{fig:ResoTime} as a function of the drift distance and visible energy. A visible energy cut ($<$50\,MeV) has been applied to avoid events with poor photon statistics\footnote{This represents $\sim5\%$ of the neutrino events with a reconstructed OpFlash.}. We obtain an almost flat bias of 2\,ns (within $\sim$1\,ns) for most drift distances and energy depositions. Note that the maximum bias in $\langle \text{X$_\text{reco}$} \rangle$ (about 15\,cm very close to the anode) represents a bias in time of $\mathcal{O}$(1\,ns). The standard deviation is also at the level of 2\,ns, except for energy depositions below $\sim$200\,MeV where the resolution is expected to be worse due to the small number of photons detected in the events, affecting the resolution in the OpFlash-time reconstruction.

The lower number of photons detected, the slower time response and the longer sampling interval (16\,ns) prevent the X-ARAPUCA system from reaching a resolution below $\mathcal{O}$(10\,ns), as the OpFlash time determination is affected by these system parameters. Future reconstruction techniques beyond those used for this work may overcome this hardware limitation in the X-ARAPUCAs.

\section{Application example: resolving the BNB bucket structure}\label{sec:BNBreconstruction}

\begin{figure*}[ht!]
\centering
     \includegraphics[width=.49\textwidth]{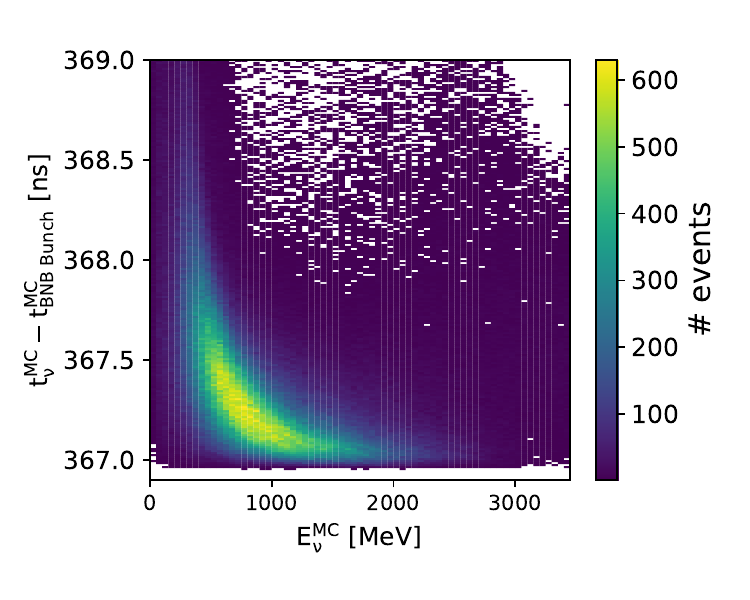}
     \includegraphics[width=.49\linewidth]{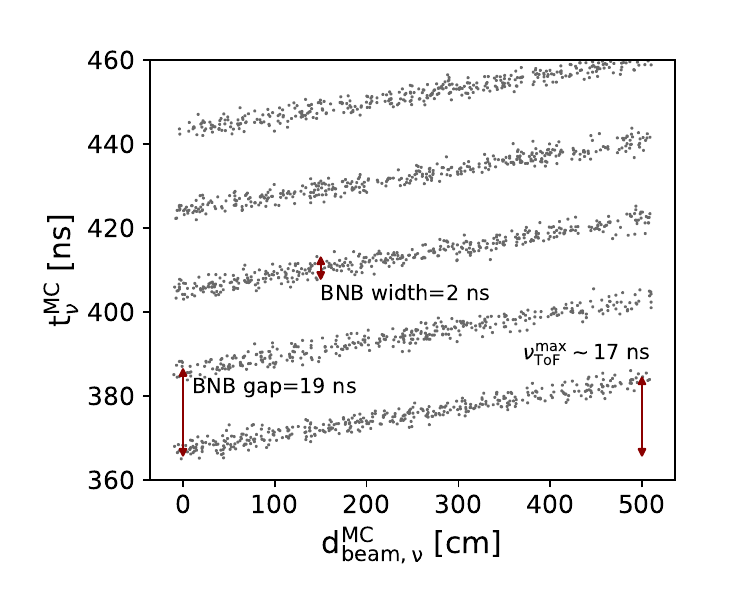}
 \caption{Left: Time delay of neutrinos arriving to the most upstream SBND wall relative to the proton bunch delivery time, as a function of their energies. The minimum delay around 367\,ns represents the case when the parent hadron decays just after the beryllium target, at 110\,m from the beginning of the active volume. Right: BNB-neutrinos ToF along the 5\,m length active volume of SBND TPC, for the first five bunches, including the simulated parent hadron decay time (left panel).} \label{fig:DelayTime}
\end{figure*}

The BNB is created by extracting protons from the Booster accelerator at Fermilab and impacting them on a beryllium target. The time structure of the delivered proton beam consists of a series of 81 bunches, each about 1.3\,ns wide and 19\,ns apart, defining a spill length of 1.6\,\textmu s~\cite{osti_1128043, RWMBeamMeasurement}.

An OpFlash in time coincidence with the expected arrival of the proton delivery spill is a strong indication for a neutrino interaction. However, cosmic rays can also interact during the proton beam delivery, resulting in background triggers. One out of $\sim$300 beam spills are expected to have an in-time cosmic muon~\cite{acciarri2015proposal}. SBND is potentially able to reduce the random cosmogenic background happening during the beam spill. As we have seen in section~\ref{sec:timeres}, the OpFlash t$_0$ gives an estimation of the neutrino interaction time 
with a resolution of the order of 2\,ns for our PMT system. This timing resolution allows us to correlate the reconstructed OpFlash with the BNB bunches, allowing SBND to develop sophisticated selection criteria and reject cosmic interactions happening between the beam bunches.

\begin{figure*}[ht!]
    \centering
    \includegraphics[width=.99\linewidth]{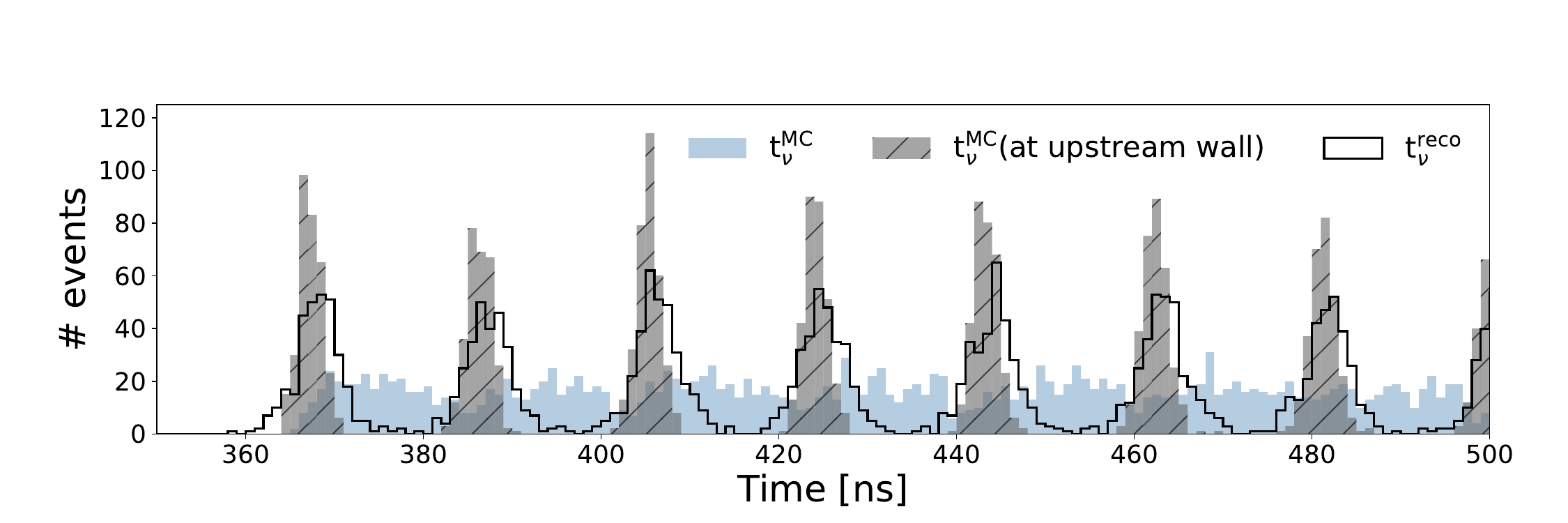}
     \caption{True neutrino arrival time distributions at the SBND upstream wall (dashed-grey) and inside the active volume (solid-blue), and reconstructed neutrino interaction time after ToF correction (black). All the times are referred to the proton on target interaction time.} \label{fig:BNBTime}
\end{figure*}

However, there are two sources of delays that cause the interaction times of neutrinos to  differ from that of the proton extraction time structure. One is due to the decay time of their parent hadrons. To quantify this effect, Figure~\ref{fig:DelayTime}-Left shows the delay of the neutrinos arriving to the frontmost wall of the detector relative to the proton bunch delivery time~\cite{PhysRevD.79.072002}. We see how the cumulative delays caused by this effect are below 2\,ns. The second source of delay is due to neutrinos interacting along the 5\,m length of SBND's active volume in the beam direction. Therefore, there will be an additional delay that depends on where the neutrino interacts in the TPC. This additional time ranges from 0 to about 17\,ns, for neutrinos interacting in the upstream or the downstream wall of the detector respectively. As can be seen in Figure~\ref{fig:DelayTime}-Right, proper reconstruction of the interaction point along the detector length allows to resolve the individual neutrino bunches.

Figure~\ref{fig:BNBTime} illustrates how the beam time structure degrades for the neutrino interaction times. This undesired effect could be minimised by correcting for the position of the neutrino interaction in the beam direction (Z). The millimetre-level particle tracking capability of the LArTPC technology will enable this correction in a very precise manner~\cite{MicroBooNE:2023ldj}. However, in this work we will use the method described in section~\ref{sec:position} to demonstrate the PDS-only performance reconstructing the Z coordinate. The result after this correction is also shown in Figure~\ref{fig:BNBTime}. This demonstrates how using only the scintillation light signals recorded by the PMTs in SBND, we are able to precisely correct the different sources of delay to the neutrino interaction time and recover the BNB time structure. 

\begin{figure}[ht!]
    \centering
    \includegraphics[width=1.\linewidth]{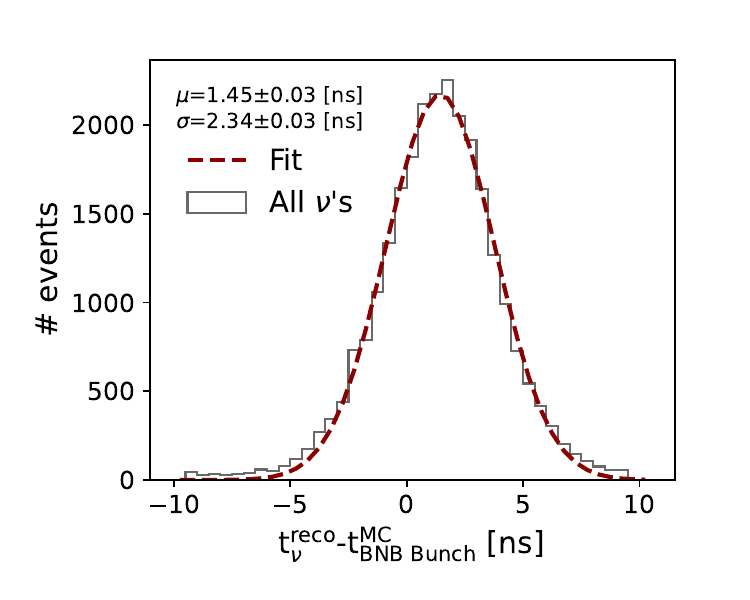}
    \caption{Neutrino interaction times of the BNB bunches merged into a single peak after applying all the corrections described in the text.}
    \label{fig:bnb_merged}
\end{figure}

Finally, to quantify the overall timing resolution we have fit with a Gaussian function all the 81 bunches merged in one single peak. As can be seen in Figure~\ref{fig:bnb_merged}, we get an average bias of 1.45\,ns with a standard deviation of 2.34\,ns. If we subtract from this width the intrinsic BNB bunch width $\langle \sigma_\text{BNB} \rangle$=1.3\,ns, we end up with a global timing resolution of 2\,ns, as expected.

\section{Conclusions}
SBND has the most advanced PDS to date installed in a neutrino LArTPC. It integrates passive elements with a high density of detectors using two different technologies: 8-inch cryogenic PMTs and a variety of new X-ARAPUCA devices. Both systems mix WLS-coated and uncoated units to be sensitive to the direct and reflected light components available in SBND. This PDS design increases the LY and makes it more uniform across the whole detector (by more than doubling the amount of light detected in the region furthest away from the plane of detection, thanks to the addition of the reflected light component), with the aim of improving its performance and extending its use in physics analysis.

The experiment employs an efficient and accurate optical simulation, overcoming the challenges that this entails mainly due to the prolific nature of liquid argon as a scintillator. The simulation model is hybrid, combining two photon propagation approaches to simulate scintillation light in the entire argon volume and for our two different light components, with high accuracy in both the number of detected photons and their arrival times. 

SBND has employed a deconvolution procedure to undo the effects of AC-coupling of our detectors. We have demonstrated that with this procedure we are able to reconstruct the number of PEs for BNB events with a resolution better than 4\% for both systems of PMTs and X-ARAPUCAs. This, together with the high LY opens the door to improving the accuracy of calorimetric energy reconstruction by incorporating the information provided by our PDS signals. Future work will explore the combination of PDS and TPC systems for this purpose.

In addition, the novel ability to distinguish between direct and reflected light components allows SBND to reconstruct the average drift coordinate where the energy depositions occur, with a resolution between $10-15$\,cm ($10-20$\,cm) for the PMT (X-ARAPUCA) system, and allows a 3D reconstruction of the average position of the interactions using only scintillation light. This possibility has, for instance, the potential to improve our signal vs background tagging, which can be challenging in near-surface detectors like SBND. It could also enable 3D-based readout of parts of the detector.

Finally, we have shown that with the reconstruction algorithms developed in this work, the SBND PMT system, which is the system used to construct trigger signals, can reconstruct the time of the events with a resolution of the order of 2\,ns. This result, together with the fact that the SBND PDS can accurately reconstruct the positions where particles interact, makes it possible to correct for the neutrinos' ToF inside the detector (from the Z-coordinate reconstruction) and the photons ToF until they are recorded (from the drift-coordinate reconstruction). This allows us to correlate the neutrino interaction times with the times of the proton delivery that generate the neutrino beam, and therefore to identify events occurring within or outside the BNB time structure. This event-by-event capability has, for example, applications in searches for long-lived massive particles as predicted in some beyond Standard Model physics models, or in rejecting cosmogenic events in coincidence with the beam.

\begin{acknowledgements}
The SBND Collaboration acknowledges the generous support of the following organisations: 
the U.S. Department of Energy, Office of Science, Office of High Energy Physics; 
the U.S. National Science Foundation; 
the Science and Technology Facilities Council (STFC), part of United Kingdom Research and Innovation, The Royal Society of the United Kingdom, and the UK Research and Innovation (UKRI) Future Leaders Fellowship; 
the Swiss National Science Foundation; 
the Spanish Ministerio de Ciencia e Innovación (MICIN/ AEI/ 10.13039/ 501100011033) under grants No PRE2019-090468, PID2019-104676GB-C31 \& C32, RYC2022-036471-I, and Comunidad de Madrid (2019-T2/TIC-13649); 
the European Union’s Horizon 2020 research and innovation programme under GA no 101004761 and the Marie Sklodowska\,-\,Curie grant agreements No 822185 and 892933; 
the São Paulo Research Foundation 1098 (FAPESP), the National Council of Scientific and Technological Development (CNPq) and Ministry of  Science, Technology \& Innovations-MCTI of Brazil. 
We acknowledge Los Alamos National Laboratory for LDRD funding. 
SBND is an experiment at the Fermi National Accelerator Laboratory (Fermilab), a U.S. Department of Energy, Office of Science, HEP User Facility. 
Fermilab is managed by Fermi Research Alliance, LLC (FRA), acting under Contract No. DE-AC02-07CH11359.
\end{acknowledgements}

\printbibliography

\end{document}